\documentclass[10pt,aps,prd,showpacs,nofootinbib,twocolumn,superscriptaddress]{revtex4-2}

\usepackage{amssymb,amsmath,amsfonts}
\usepackage{xcolor}
\usepackage[pdftex]{graphicx}
\usepackage[english]{babel}
\usepackage{amsmath,amssymb}
\usepackage{multirow}
\usepackage[super]{nth}
\usepackage{txfonts}
\usepackage[normalem]{ulem}
\usepackage[percent]{overpic}

\usepackage{lipsum}
\usepackage{hyperref}

\usepackage{stackengine}
\newcommand{\raisedddot}[1]{\stackon[1.5pt]{${#1}$}{..}}

\begin{document}

\title{Universal Relations with Dynamical Tides}
\author{Jayana A. Saes}
\email{jayanaa2@illinois.edu}
\affiliation{Illinois Center for Advanced Studies of the Universe, Department of Physics, University of Illinois Urbana-Champaign, Urbana, IL 61801, USA}
\author{Abhishek Hegade K. R.}
\email{ah4278@princeton.edu}
\affiliation{Illinois Center for Advanced Studies of the Universe, Department of Physics, University of Illinois Urbana-Champaign, Urbana, IL 61801, USA}
\affiliation{Princeton Gravity Initiative, Princeton University, Princeton, NJ 08544, USA}
\author{Nicol\'as Yunes}
\email{nyunes@illinois.edu}
\affiliation{Illinois Center for Advanced Studies of the Universe, Department of Physics, University of Illinois Urbana-Champaign, Urbana, IL 61801, USA}

\date{\today}

\begin{abstract}
Observations of neutron stars and the precise measurement of their macroscopic properties have provided valuable insights into fundamental physics, both by constraining the behavior of nuclear matter under extreme conditions and by enabling tests of general relativity in the strong-field regime. In this context, equation-of-state–insensitive or ``quasi-universal'' relations between key observables, such as the compactness, dimensionless static tidal deformability, and moment of inertia, play a crucial role in connecting different measurable observables while minimizing uncertainties due to the yet unknown equation-of-state. In this work, we identify new quasi-universal relations between the static, dimensionless tidal deformability ($\Lambda^{(0)}$) and its leading-order dynamical correction ($\Lambda^{(2)}$), as well as between $\Lambda^{(0)}$ and a combination of these parameters ($\sqrt{ \Lambda^{(0)}/\Lambda^{(2)}}\equiv M\omega_*$), obtained from the small-frequency expansion of the relativistic tidal response. We test these relations across a representative set of 59 equations of state, finding that the equation-of-state dependence does not exceed $\sim$5\% for the $\Lambda^{(0)}$--$\Lambda^{(2)}$ relation and $\sim2.8\%$ for the $\Lambda^{(0)}$--$M\omega_*$ relation. This indicates a high degree of universality and offers a simplified framework for incorporating dynamical tidal effects into gravitational-wave modeling. Furthermore, we compare the dynamical tidal response against different recent strategies (a Taylor expansion and a one-mode approximation) to model the dynamical tide. We find that both models are capable of capturing the frequency-dependent behavior of the dynamical tidal deformability, with the one-mode approximation agreeing better with the dynamical response than the Taylor expansion in most of the parameter space.
\end{abstract}
\maketitle
\allowdisplaybreaks[4] 
\section{Introduction}

Neutron stars (NSs) are highly compact astrophysical objects inside which the matter density is extremely high. For a NS with a characteristic mass of $1.4 $ M$_{\odot}$ and radius of about $10$ km, its central (maximum) density can reach several times nuclear saturation density, $n_\text{sat} = 0.16$ fm$^{-3}$. At such high densities, nucleons cannot exist in isolation, potentially allowing for new states of matter. As such, NSs serve as important laboratories for the study of the equation of state (EOS) that governs matter at high densities~\cite{Markakis:2009mzp,Miller:2019nzo}, as well as for probing general relativity itself (see~\cite{Silva:2024cit} for a review).

Over the past decade, precise astrophysical observations of NSs have attempted to provide insight into the EOS of supranuclear matter. This is possible because the observable properties of NSs, such as their mass, radius, and tidal deformability, are highly dependent on the microphysics of the star~\cite{LIGOScientific:2018cki,Read:2009yp}. Experimental observations from pulse-profiles of X-ray data by the Neutron-Star-Interior Composition Explorer (NICER)~\cite{Miller:2019cac,Miller:2021qha}, gravitational-wave (GW) observations of the late inspiral of binary NSs (BNSs) by the LIGO/Virgo/KAGRA collaboration~\cite{LIGOScientific:2017vwq,LIGOScientific:2018cki}, and the radio observations of pulsars~\cite{Kramer:2006nb,Weisberg:2010zz, Fonseca:2014qla} have already significantly improved our knowledge of the physics at play in NS environments. 

During the late stages of a BNS inspiral, each NS in the binary becomes tidally deformed by its companion’s gravitational field. This tidal response depends sensitively on the internal structure of the star, encoded in the EOS, and leaves a characteristic signature on the GWs emitted~~\cite{Flanagan:2007ix,Hinderer:2009ca,GuerraChaves:2019foa}, becoming more significant at GW frequencies of around $f_\text{GW} \sim 400$ Hz and higher~\cite{Harry:2018hke}. The modeling of this response, characterized by its tidal deformability, is therefore essential for both GW data analysis and constraining the dense matter EOS. 

Over the past decades, several frameworks have been developed to describe the tidal deformability of NSs in many contexts. Early work focused on understanding the tidal response in the Newtonian regime~\cite{Lai:1993di,1994ApJ...426..688R,Kokkotas:1995xe,Ho:1998hq}, particularly on how it connects to the NS oscillation modes. In this context, the oscillation modes describe the characteristic ways in which the fluid within the star oscillates when subjected to perturbations. The fundamental ($f$-) mode corresponds to the global, pressure-driven oscillation of the star, and typically occurs at kHz frequencies. In contrast, the gravity ($g-$) modes arise from buoyancy forces associated with thermal or composition gradients, or from composition discontinuities within the stellar interior~\cite{Finn:1987zoj,Kuan:2021jmk,Kuan:2021sin,Kuan:2022bhu}. These studies~\cite{Lai:1993di,1994ApJ...426..688R,Kokkotas:1995xe,Ho:1998hq,Yu_2016,Yu:2024uxt,Gao:2025aqo} allowed for a first understanding of how the resonant excitations of the $f-$ and the $g-$modes of NSs can influence the tidal response. 
More recently, studies have also tried to understand how non-linear mode interactions can lead to resonance locking and non-linear tidal coupling~\cite{Kwon:2025zbc,Kwon:2024zyg,Yu:2022fzw}.
The assumption of a Newtonian treatment of the tidal response, however, is not valid for NSs, which are highly relativistic.

Later work extended these analyses into the relativistic regime under the approximation of a static or adiabatic response~\cite{Hinderer:2007mb, Binnington:2009bb,Damour:2009vw}, where one assumes that the tidal field varies slowly compared to the timescales of the NS oscillation modes, or equivalently, that the orbital frequency of the binary is small compared to the mode frequency. In this scenario, the tidal deformability is treated as a single constant number, which depends only on the EOS and the central density of the star. In reality, when the orbital frequency is high enough, resonances can occur between the tidal driving force and the internal modes of the star. The assumption of a time-independent (adiabatic) response breaks down in this case, and the tidal deformability acquires a frequency dependence. This gives rise to the so-called \textit{dynamical tides}, where the tidal response increases near resonance and depends sensitively on the coupling between the tidal field and the star’s oscillation. Subsequent studies done using an effective-one-body (EOB) approach~\cite{Hinderer:2016eia,Steinhoff:2016rfi}, introduced the concept of these dynamical tides into waveform modeling, effectively parameterizing the frequency-dependent tidal effects in a form that is suitable for GW data analysis. However, the EOB implementation, while phenomenologically successful, requires an underlying basis from relativistic perturbation theory to completely understand the systematics and to incorporate nonlinear effects.

More recently, several studies have tried to approach the problem of dynamical tidal interactions in general relativity using perturbation theory, matched asymptotic expansions and effective field theory tools~\cite{Chakrabarti:2013lua,Poisson:2020vap,Pitre:2023xsr,Pitre:2025qdf,HegadeKR:2025qwj,Andersson:2025iyd,Miao:2025utd}. In particular, Pitre \& Poisson~\cite{Pitre:2023xsr} developed a framework to compute the frequency-dependent tidal response directly from linear perturbations of the Einstein equations, therefore not relying on Newtonian analogies. Their method provided the first systematic relativistic treatment of dynamical tides and established the connection between the frequency-dependent Love number and the internal stellar structure. However, the approach presented in~\cite{Pitre:2023xsr} is performed in the small-frequency expansion, both internally and externally; such expansion becomes unreliable in the presence of low-frequency $g$-modes and during the late inspiral. 

Hegade \textit{et al.}~\cite{HegadeKR:2024agt} moved beyond the small frequency approximation by formulating the interior problem in terms of a system of coupled ``master equations'' that are solved directly in the frequency domain, capturing the full frequency-dependent fluid dynamics, while the exterior problem is treated analogously to that of~\cite{Pitre:2023xsr}, but with the introduction of a resummation scheme in the frequency domain. This formulation naturally incorporates higher-order, time-dependent, fluid dynamical effects and accurately captures resonance phenomena associated with $g$-modes, which are inaccessible to the perturbative approach of Pitre \& Poisson.
More recently, it was shown that the resummation approach can provide a set of complete eigenfunctions that can be used to obtain mode-sum approximations used in EOB models~\cite{HegadeKR:2025qwj,Andersson:2025iyd}.
The mode-sum approach has also been applied, using a phenomenological approach, to capture $g$-mode resonances in~\cite{Andersson:2025iyd}.
With the theoretical basis for dynamical tidal interactions established, it is now time to incorporate this relativistically consistent formalism into waveform modeling and compare it with EOB models~\cite{Abac_2024}. 

A valuable tool that aids the measurement and parameter estimation of the static tidal deformability is the existence of quasi-universal or EOS-insensitive relations. For instance, the I-Love-Q relations~~\cite{Yagi:2013awa} link the dimensionless moment of inertia to the tidal deformability and to the quadrupole moment of a NS, providing a way to break degeneracies in NICER and GW parameter estimation. Similarly, the binary-Love relations~\cite{Yagi:2015pkc, Yagi:2016qmr} connect the dimensionless tidal deformabilities of the two NSs in a binary, allowing individual (static) tidal deformabilities to be extracted from GW observations~\cite{LIGOScientific:2018cki}. Another important example is the $f$-mode - Love relation~\cite{Chan:2014kua,Zhao:2022tcw}, which establishes a tight relation between the frequency associated with the star's $f$-mode oscillation frequency and its (static) tidal deformability, which has been implemented in EOB models~\cite{Abac_2024}. 

In this work, we use the relativistic approach of~\cite{Pitre:2023xsr,HegadeKR:2024agt} to answer the following questions:
\begin{itemize}
    \item [1.] How do the low-frequency expansion of~\cite{Pitre:2023xsr} and the EOB mode-sum~\cite{Steinhoff:2016rfi} compare to the dynamical approach of~\cite{HegadeKR:2024agt}? In particular, over what frequency and compactness range do these approximations begin to deviate from the dynamical calculation?
    \item [2.]  What quasi-universal relations exist among the coefficients appearing in the low-frequency expansion of the dynamical tidal response function, and to what extent are these relations insensitive to the NS EOS?
\end{itemize}
To perform the comparison, we adopt two different approximations to the dynamical tidal response. 
First, we use the approach of~\cite{Pitre:2023xsr} and consider a Taylor series expansion up to second order in small frequencies ($M\omega \ll 1$) for the dynamical tidal response function, $\Lambda^{\rm Tay} (\omega)= \Lambda^{(0)} + \Lambda^{(2)} (M\omega)^2$, where $\Lambda^{(0)}$ corresponds to the static tidal deformability, $\Lambda^{(2)}$ corresponds to the first correction due to the frequency dependence, $M$ is the NS mass, and $\omega=2 \pi f_{\rm GW}$ is the angular GW frequency. 
Second, we use a one-mode representation, used commonly in EOB models, $\Lambda^{{\rm{OM}}}(\omega)= \Lambda^{(0)} (1-\omega^2/\omega_*^2)^{-1}$, where $\omega_*$ is the effective $f-$mode frequency of the NS, obtained using $\omega_* = M^{-1} \sqrt{ \Lambda^{(0)}/\Lambda^{(2)}}$.

We study these approximations by directly comparing them to the dynamical tidal response $\Lambda^{\rm Dyn}(\omega)$ up to linear frequencies of 1.5 kHz. Our results show that, for linear frequencies $\lesssim 750$ Hz, both models reproduce the dynamical result within $5\%$. At higher frequencies, $\Lambda^{\rm OM} (\omega)$ shows better agreement; the relative difference of the OM approximation remains below $10\%$ up until $f_{\rm GW} \gtrsim 1400$ Hz, while the Taylor approximation exceeds $10\%$ at around $900$ Hz.

These findings suggest that both the low-frequency model and the one-mode model are useful representations of the frequency-dependent behavior of the tidal response of a NS. While they provide an accurate description of the tidal response, they both still involve two quantities ($\Lambda^{(0)}$ and $\Lambda^{(2)}$), and therefore, they do not simplify the parameter estimation problem on their own. To overcome this, we identify and characterize a novel quasi-universal relation between $\Lambda^{(0)}$ and $\Lambda^{(2)}$, as well as between  $\Lambda^{(0)}$ and a combination of these two quantities $ \sqrt{ \Lambda^{(0)}/\Lambda^{(2)}}= M\omega_*$. We present and explore the robustness of these quasi-universal relations across a sample of 59 EOSs, including both parametrized, piecewise-polytropic models~\cite{Read:2008iy} and phenomenologically-motivated EOSs that incorporate nontrivial microphysical features~\cite{Tan:2021ahl}. Our results show that the EOS-sensitivity of the relation is remarkably small, $\lesssim 5\%$ for the $\Lambda^{(0)}$--$\Lambda^{(2)}$ relation and $\lesssim 2.8\%$ for the $\Lambda^{(0)}$--$M\omega_*$ relation. These quasi-universal relations enable the dynamical tidal response to be modeled with effectively a single independent parameter. Doing so offers a practical route for incorporating frequency-dependent tidal effects into GW waveforms without introducing additional degrees of freedom, and thereby, improving the efficiency and robustness of parameter estimation. 

The remainder of this work presents the details of the results described above and is organized as follows. In Sec.~\ref{sec:dynamicaltides}, we outline the derivation and details that go into calculating $\Lambda^{(2)}$, following the procedure presented in~\cite{Pitre:2023xsr, HegadeKR:2024agt}. We introduce the Taylor series approximation $\Lambda^{\rm Tay}(\omega)$, as well the one-mode parametrization $\Lambda^{\rm OM}(\omega)$. In Sec.~\ref{sec:validity}, we assess the regime of validity of these two models and quantify their agreement with the dynamical tide. Finally, in Sec.~\ref{sec:universalrelation}, we present the resulting universal relations for both $\Lambda^{(0)}-\Lambda^{(2)}$ and $\Lambda^{(0)}-M\omega_*$, demonstrate their degree of universality, and discuss the potential implications for NS parameter inference. Throughout this paper, we use geometric units in which $G=1=c$. 

\section{Calculation of Dynamical Tides}
\label{sec:dynamicaltides}
The calculation of the static tidal deformability is described in detail in~\cite{Hinderer:2007mb,Binnington:2009bb,Damour:2009vw}.
One begins with a spherically-symmetric background NS, obtained from the solution to the Tolman–Oppenheimer–Volkoff (TOV) equations. Linear and static perturbations due to a static quadrupolar external tidal field are then introduced, which affect both the metric and the fluid. The linearized Einstein equations and the stress-energy conservation equations can be organized into a single master equation~\cite{Hinderer:2007mb,Binnington:2009bb,Damour:2009vw}.
This master equation is solved by imposing regularity at the origin and continuity and differentiability at the surface of the star. The value of the perturbed metric tensor at the surface of the star can be related to the static dimensionless tidal deformability $\Lambda^{(0)}$.

In reality, however, when two NSs orbit each other, the tidal field is not strictly static, but rather it varies slowly in time. In the early stages of the inspiral, the orbital period is long compared to the star’s internal dynamical timescales, so the static approximation is excellent. However, as the stars approach their late inspiral stage and the orbital frequency increases, the tidal field can vary on timescales comparable to the natural oscillation frequencies of the star, such as the $f-$mode frequency~\cite{Steinhoff:2016rfi,Ma:2020rak,Poisson:2020vap,Pitre:2023xsr,HegadeKR:2024agt,Yu:2024uxt,Yu:2025ptm}. In this regime, the star’s response becomes dynamic, and consequently, the tidal deformability becomes frequency dependent.

The methodology to obtain the dynamical tidal deformability in general relativity is described in detail in Ref.~\cite{HegadeKR:2024agt}. In this approach, one considers the time-dependent, non-radial, first-order perturbations of a static and spherically symmetric NS background. The time dependence is assumed to be harmonic, proportional to $e^{i\omega t}$.
Similar to the static case, the linearized Einstein equations can be simplified to a set of coupled, ordinary differential equations of the relevant perturbation functions, which depend explicitly on the frequency $\omega$. These include the fluid displacement functions $\{W(r,\omega), V(r,\omega)\}$ and the metric perturbation function $H(r,\omega)$, which carries information about the tidal deformation of the star.
Schematically, the master equations are
\begin{align}
\label{eq:masterequations}
&W' = \alpha_{W,0} H + \alpha_{W,1} W + \alpha_{W,2} V + \alpha_{W,3} H' \nonumber \\
&V' = \alpha_{V,0} H + \alpha_{V,1} W + \alpha_{V,2} V + \alpha_{V,3} H' \\
&H'' = \alpha_{H,0} H + \alpha_{H,1} W + \alpha_{H,2} V + \alpha_{H,3} H', \nonumber
\end{align}
where a prime indicates derivative with respect to the radial coordinates, and where the coefficients $\{\mathbf{\alpha}\}$ depend on the equilibrium structure of the NS, such as the mass distribution and the pressure profile of the background star, as well as the frequency $\omega$ and the radial coordinate $r$. The explicit expressions for the coefficients $\{\mathbf{\alpha}\}$ are available in~\cite{HegadeKR:2024agt}.

To obtain the dynamical tidal response, the master equations are integrated from the center of the star to the surface, where the interior solution is matched to the exterior solution by requiring continuity and differentiability across the surface. Outside the star, the fluid perturbations vanish, leading to a single, second-order differential equation for $H$, which can be solved analytically using a small frequency expansion. From the asymptotic behavior of $H$, the induced quadrupole moment and the external tidal field can be obtained, which then allows for the calculation of the frequency-dependent, \textit{dimensionless} tidal Love number $k_2(\omega)$, and subsequently, the dynamical tidal deformability
\begin{equation}
\label{eq:dynamicaltidal}
    \Lambda^\text{Dyn}(\omega) \equiv \frac{2 k_2(\omega)}{3 C^5}\,,
\end{equation}

In this work, however, we are interested in the small-frequency expansion of the tidal deformability~\cite{Pitre:2023xsr} and in how the EOS affects the relation between the two leading-order terms in this expansion. Let us then expand the dimensionless dynamical Love number $k_2(\omega)$ in powers of the angular gravitational-wave frequency $\omega$, so that we can separate the leading-order static contribution from its first dynamical correction. Doing so, we find
\cite{Pitre:2023xsr,HegadeKR:2024agt}
\begin{equation}
\label{eq:expansionforkomega}
k_2 (\omega) \approx k_{2}^{(0)}+ \omega^2 k_{2}^{(2)} \equiv k_2^{\rm Tay}(\omega),
\end{equation}
which is valid to $\mathcal{O}(\omega ^3)$. This equation defines the Taylor expansion of the dynamical Love number, where $k_{2}^{(0)}$ represents the $l=2$ (quadrupolar) static tidal deformability and its first dynamical correction is given by $ k_{2}^{(2)}$ (both of which are frequency independent). Note here that $k_2^{(0)}$ is dimensionless, while $k_2^{(2)}$ has units of mass squared.

To proceed, we perform a small-frequency expansion of the perturbation functions and the coefficients of these equations, as presented in~\cite{Pitre:2023xsr}. Specifically, the functions $W$, $V$, and $H$, and the coefficients $\{\alpha\}$ are expanded in powers of $\omega$. The general form of this expansion for a function $X(r,\omega)$ is 
\begin{equation}
\label{eq:expansionomega}
X(r, \omega) = X(r)^{(0)} + \omega^2 X^{(2)}(r) + \mathcal{O}(\omega^3),
\end{equation}
where $X^{(0)}(r)$ corresponds to the static (i.e., frequency-independent) limit, while $X^{(2)}(r)$ represents the first dynamical correction, both of which are $\omega$ independent.

By substituting Eq.~\eqref{eq:expansionomega} for the perturbation functions and for the coefficients $X=\{W^{(0)},V^{(0)},H,\{\alpha\}\}$ into the system of master equations [Eqs.~\eqref{eq:masterequations}], and performing a series expansion to second order in the frequency, we obtain a frequency-independent set of differential equations for the relevant functions $\{H^{(0)}(r),W^{(0)}(r),V^{(0)}(r),H^{(2)}(r)\}$. The resulting system can be written schematically as
\begin{align}
\label{eq:masterexp}
H^{(0)''} &= \alpha_{H,0}^{(0)} H^{(0)}  + \alpha_{H,3}^{(0)} H^{(0)'} ,\nonumber\\
W^{(0)'} &= \alpha_{W,0}^{(0)} H^{(0)} + \alpha_{W,1}^{(0)} W^{(0)} + \alpha_{W,2}^{(0)} V^{(0)} + \alpha_{W,3}^{(0)} H^{(0)'} , \\
V^{(0)'}  &= \alpha_{V,0}^{(0)} H^{(0)} + \alpha_{V,1}^{(0)} W^{(0)} + \alpha_{V,2}^{(0)} V^{(0)} + \alpha_{V,3}^{(0)} H^{(0)'} ,\nonumber \\
H^{(2)''} &= \alpha_{H,0}^{(2)} H^{(0)} + \alpha_{H,1}^{(2)} W^{(0)} + \alpha_{H,2}^{(2)} V^{(0)} +  \alpha_{H,3}^{(2)} H^{(0)'}+\nonumber\\& \alpha_{H,0}^{(0)} H^{(2)}  + \alpha_{H,3}^{(0)} H^{(2)'},\nonumber
\end{align}
where all the coefficients $\{\alpha\}$ depend on the equilibrium background functions and the radial coordinate; their explicit expressions can be found in Appendix~\ref{Ap:Coefs}. The first three equations correspond to the static part of the expansion, whereas the last equation carries the information that will enter at $\mathcal{O}(\omega^2)$ in the expansion. 

To solve the expanded set of master equations [Eq.~\eqref{eq:masterexp}], appropriate boundary conditions must be imposed both at the center and at the surface of the star. These conditions are determined by enforcing regularity of the perturbation functions at the origin and matching conditions at the stellar surface. The procedure to obtain these boundary conditions is described in detail in Appendix~\ref{Ap:BCs}, along with their expressions. Once the boundary conditions are established, the system can be integrated sequentially inside the star.

The procedure to solve the coupled set of equations above [Eq.~\eqref{eq:masterexp}] begins with solving the decoupled, second-order differential equation for the static metric perturbation $H^{(0)}$. Solving for this equation in the interior of the star yields the radial profile of $H^{(0)}(r)$. Matching with the exterior solution, which can be found analytically, yields the static Love number $k_{2}^{(0)}$. We then move on to solve the coupled, first-order equations for the fluid perturbation variables $W^{(0)}$ and $V^{(0)}$, which describe the internal displacement of the fluid elements. These solutions depend on the previously obtained profile $H^{(0)}(r)$ and provide the necessary input for the last equation. Lastly, we use the functions $H^{(0)}(r)$, $W^{(0)}(r)$, and $V^{(0)}(r)$ as source terms to solve for the inhomogeneous differential equation for $H^{(2)}(r)$. Matching the interior and exterior solutions of this equation at the surface yields ${k_{2}^{(2)}}$. The exterior solutions for $H^{(0)}(r)$ and $H^{(2)}(r)$ can be found in Appendix~\ref{Ap:BCs}, and the full details of the matching procedures and integration are described in Appendix~\ref{Ap:Matching}. 

After obtaining $k_{2}^{(2)}$, we now normalize this quantity so that we obtain a dimensionless Love number. From Eq.~\eqref{eq:expansionforkomega}, the coefficient $k_{2}^{(2)}$ has units of mass squared. To obtain a dimensionless quantity, and to make contact with Pitre \& Poisson (see Eq.~(1.9) in~\cite{Pitre:2023xsr}), we define
\begin{equation}
    \raisedddot{k}_{2} \equiv  \frac{M}{R^3} \left.\frac{dk_2(\omega)}{d\omega^2}\right|_{\omega =0} = \frac{M}{R^3} \frac{dk_2^{\rm Tay}(\omega)}{d\omega^2} = C^3 \frac{k_{2}^{(2)}}{ M^2},
\end{equation}
where $M$ and $R$ are the total (gravitational) mass and radius of the star, while recall that $C = M/R$ is the stellar compactness. With this normalization, the expansion in Eq.~\eqref{eq:expansionforkomega} is re-expressed as
\begin{equation}
\label{eq:dimensionlessexpk}
    k_2^{\rm Tay}(\omega) = k_{2}^{(0)}+ (M \omega)^2 \frac{\raisedddot{k}_{2}}{ C^3}.
\end{equation} 
Substituting Eq.~\eqref{eq:dimensionlessexpk} in Eq.~\eqref{eq:dynamicaltidal} we obtain a small frequency expansion of the tidal deformability
\begin{equation}
\label{eq:taylorlambda}
\Lambda (\omega) \approx \Lambda^{(0)} + \Lambda^{(2)} (M \omega)^2= \Lambda ^\text{Tay}(\omega)\,.
\end{equation} 
We now notice that $\Lambda^{\rm Tay}(\omega) = 2 k_2^{\rm Tay}(\omega)/(3 C^5)$, which then leads to
\begin{equation}
\label{eq:approxdynamical}
    \Lambda^{(0)}+ \Lambda^{(2)} (M \omega)^2 = \frac{2}{3 C^5} \left[k_{2}^{(0)}+ (M \omega)^2 \frac{\raisedddot{k}_{2}}{ C^3}\right]
\end{equation}
which, in turn, yields
\begin{equation}
\label{eq:lambdadef}
      \Lambda^{(0)} \equiv \frac{2k_{2}^{(0)}}{3 C^5}  , \quad      \Lambda^{(2)} \equiv \frac{2\raisedddot{k}_{2}}{3 C^8}.
\end{equation}
These two quantities will play a key role in the universal relation we investigate later. 

To introduce the effective frequency $\omega_*$, we note that the dynamical Love number is usually represented through a single effective-mode ~\cite{Pitre:2023xsr,Steinhoff:2016rfi}:
\begin{equation}
k_2(\omega) \approx \frac{k_{2}^{(0)}}{1 - \omega^2 / \omega_*^2} \equiv k_2^{\rm OM}(\omega).
\end{equation}
Here, $\omega_*$ denotes approximately the $f$-mode frequency of the star. Although $\omega_*$ does not correspond to a true normal mode of oscillation of the star, this expression provides a way to capture the frequency dependence of the Love number within a one-mode framework. The use of this approximation is widespread in the EOB modeling community, where it serves as a tool for incorporating dynamical tidal effects into GW waveform models. This representation, however, is only an approximation, since the true dynamical response of the star involves a spectrum of oscillation modes and cannot be modeled in this way. We later compare this approximation to the more robust calculation performed in~\cite{HegadeKR:2024agt}. 

Using Eq.~\eqref{eq:dynamicaltidal}, we have that the dynamical tidal deformability can be expressed as 
\begin{equation}
\label{eq:lambdaapproxonemode}
    \Lambda^{\rm OM}(\omega) =  \frac{\Lambda^{(0)}}{1-\omega^2/\omega_*^2},
\end{equation}
and expanding for $\omega^2/\omega_*^2<<1$ yields:
\begin{equation}
    \Lambda^{\rm OM}(\omega) \approx \Lambda^{(0)} \left(1+\frac{\omega^2}{\omega_*^2}\right).
\end{equation}
By comparing this result with Eq.~\eqref{eq:approxdynamical}, we can directly identify
\begin{equation}
    \omega_*^2 = \frac{1}{M^2}\frac{\Lambda^{(0)}}{\Lambda^{(2)}} = \frac{C^3}{M^2}\frac{k_{2}^{(0)}}{\raisedddot{k}_{2}},
    \label{eq:omega*}
\end{equation}
with $f_* \equiv \omega_*/2\pi \sim 1.8$ kHz for a $M=1.4$ M$_{\odot}$ star for the SLy EOS, while the $f-$mode for the same star has a linear frequency of around 1.7 kHz.
This expression provides a simple physical interpretation for $\omega_*$: it represents the effective frequency scale governing the leading-order dynamical correction to the static tidal deformability. Alternatively, the time scale associated with the frequency dependence of the tidal response is simply
\begin{align}
    T_* = \frac{2 \pi }{\omega_*} = {2 \pi \,M}\sqrt{\frac{\Lambda^{(2)}}{\Lambda^{(0)}}},
\end{align}
which is on the order of 0.6ms for a $M=1.4$ M$_{\odot}$ star with the SLy EOS.

\section{Regime of Validity of the small frequency expansion}
\label{sec:validity}
 
\begin{figure}[!ht]
    \centering
    \includegraphics[width=\linewidth]{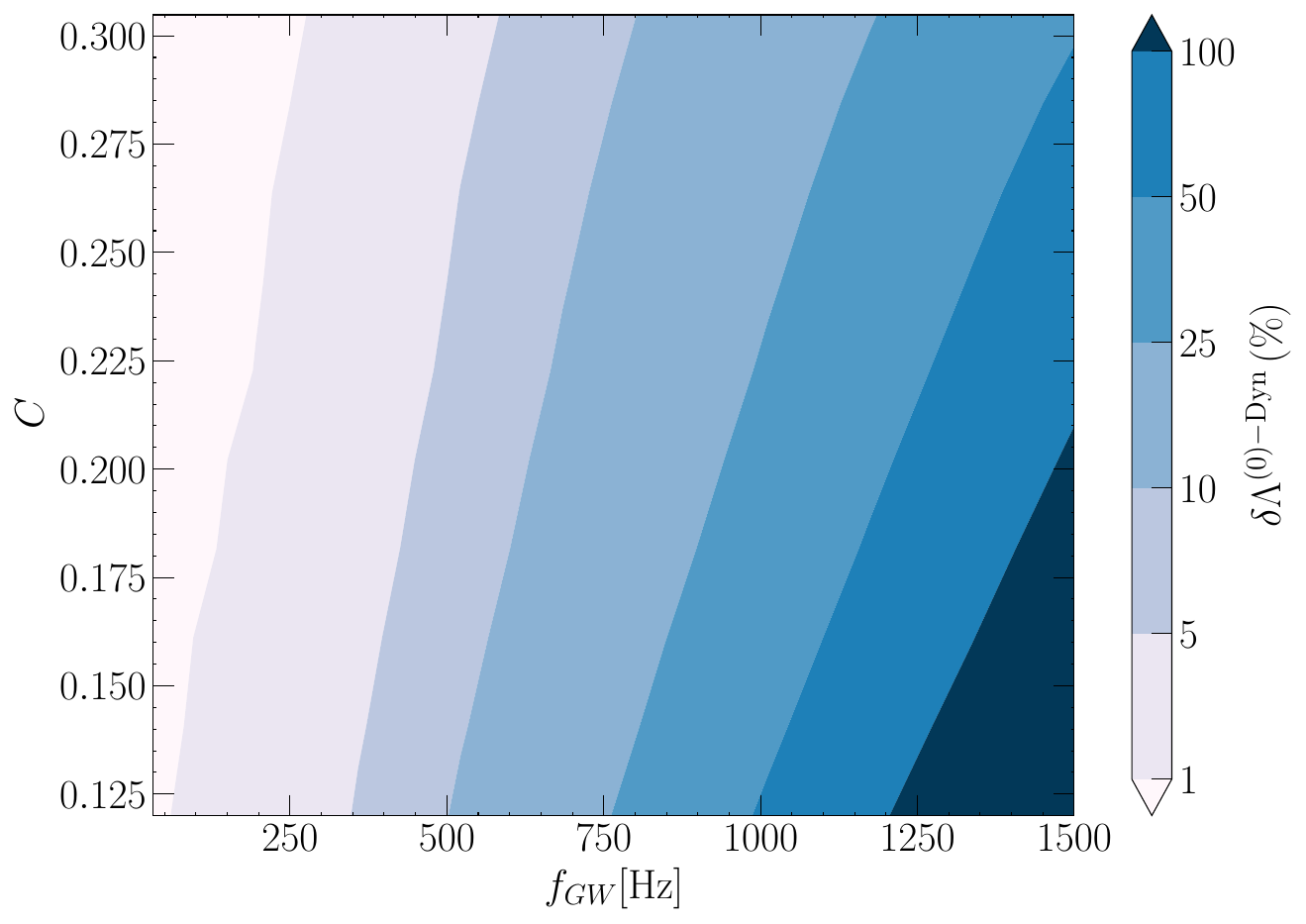}
    \caption{Fractional difference between the dynamical tide and the static tide for different values of the compactness and the GW frequency. As frequency increases, so does the contribution of the dynamical part.}
    \label{fig:lambdatruelambda}
\end{figure}

With $\Lambda^{(2)}$ in hand, direct comparisons between the static, the small-frequency expansion, and the dynamical tidal deformabilities can be made. Let us begin by assessing how the static tidal deformability compares to the dynamical one. In particular, we can quantify this by defining
\begin{equation}
   \delta \Lambda^{\rm (0)-Dyn} \equiv \Bigg |  1 -\frac{\Lambda^\text{(0)}}{\Lambda^{\rm Dyn}(\omega)}\Bigg | 
\end{equation}
for a particular EOS, across varying frequencies and compactness. The resulting behavior is presented in Fig.~\ref{fig:lambdatruelambda} for the SLy EOS. As expected, the fractional correction grows with increasing frequency. We see that the frequency dependence of the tidal deformability can modify the static tidal deformability by over 100\% for low compactness stars, and typically by over 50\% for frequency significantly below contact, which is at $f_{\rm GW}\sim$2 kHz for a 1.4 M$_{\odot}$ star. 

On the other hand, the validity of the series expansion at small frequencies can be evaluated by comparing $\Lambda^\text{Tay}(\omega)$ with the dynamical tidal deformability $\Lambda^\text{Dyn}(\omega)$, obtained from the full numerical solution of the perturbation equations presented in~\cite{HegadeKR:2024agt}. This provides a direct measure of how the Taylor expansion captures the behavior of the dynamical tidal deformability and the frequency regime in which the approximation is valid. In addition, we evaluate the performance of the one-mode approximation, $\Lambda^{\rm OM}(\omega)$ defined in Eq.~\eqref{eq:lambdaapproxonemode}, to determine how closely it reproduces the dynamical response. 

\begin{figure}[!ht]
    \centering
    \includegraphics[width=1.0\linewidth]{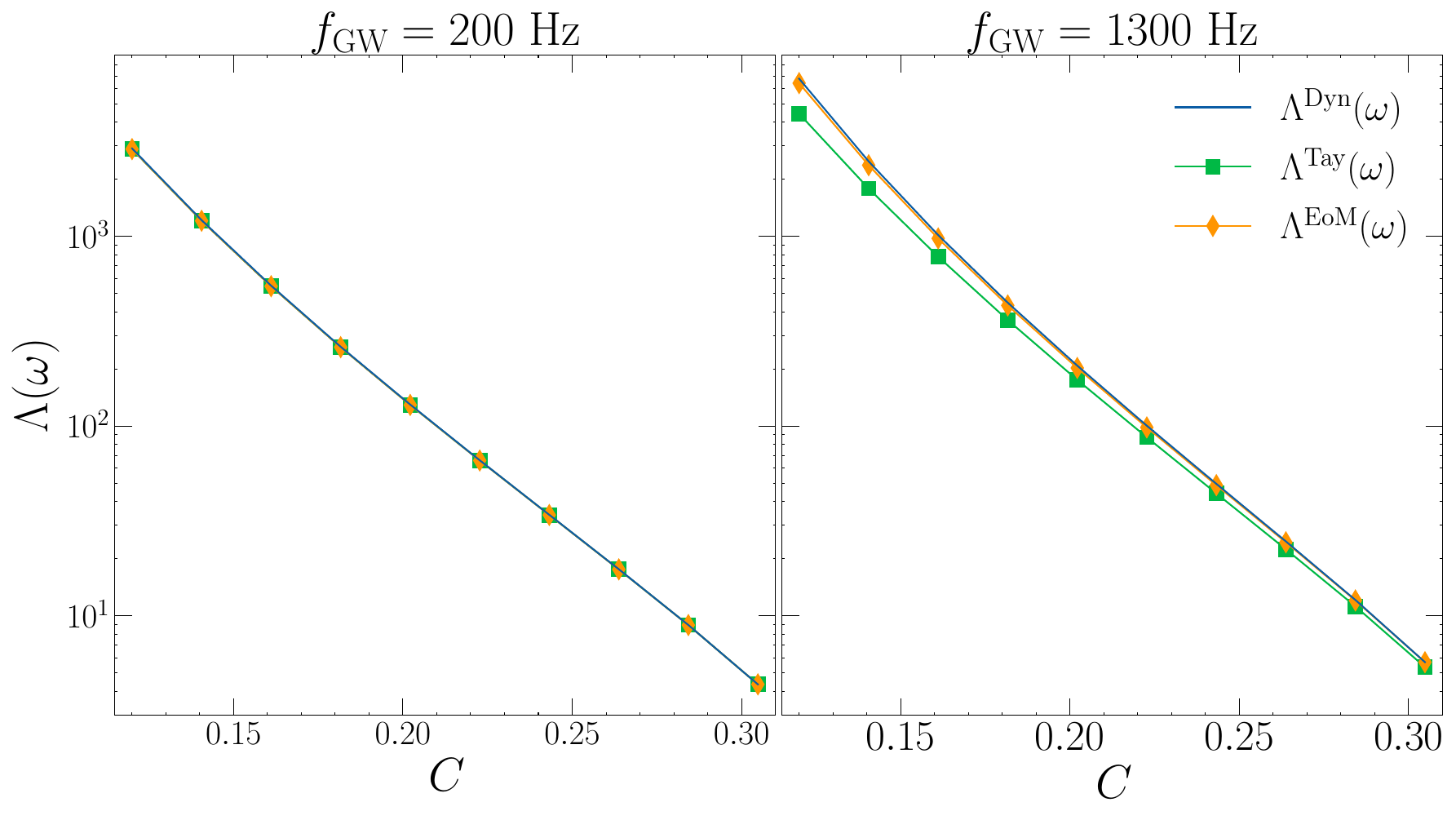}
    \caption{Frequency-dependent tidal deformability for the SLy EOS. In both panels, the dynamical tide is represented by the solid blue line; the series approximation is represented by the green line with square markers, and the one-mode approximation is represented by the orange line with diamond markers. The left panel shows the tidal deformability for $f_\text{GW}=200  $ Hz while the right panel shows it for $f_\text{GW}=1300  $ Hz.}
    \label{fig:lambdas_200_1300}
\end{figure}

To begin understanding the differences in the tidal response predicted by the three approaches, we first examine how the tidal deformability $\Lambda(\omega)$ for all three approaches varies with the stellar compactness $C$ for several fixed values of the frequency $\omega$, focusing on a representative EOS (SLy). At small frequencies, such as at $f_\text{GW} = 200$Hz, the three methods should produce very similar results, and this is indeed what we observe in the left panel of Fig.~\ref{fig:lambdas_200_1300}: the three lines lie on top of each other. At higher values of frequency, however, such as $f_\text{GW} = 1300 $Hz, the three methods start to differ from one another, highlighting the need for more careful modeling in this regime.

\begin{figure*}[htb]

   \includegraphics[width=0.48\linewidth]{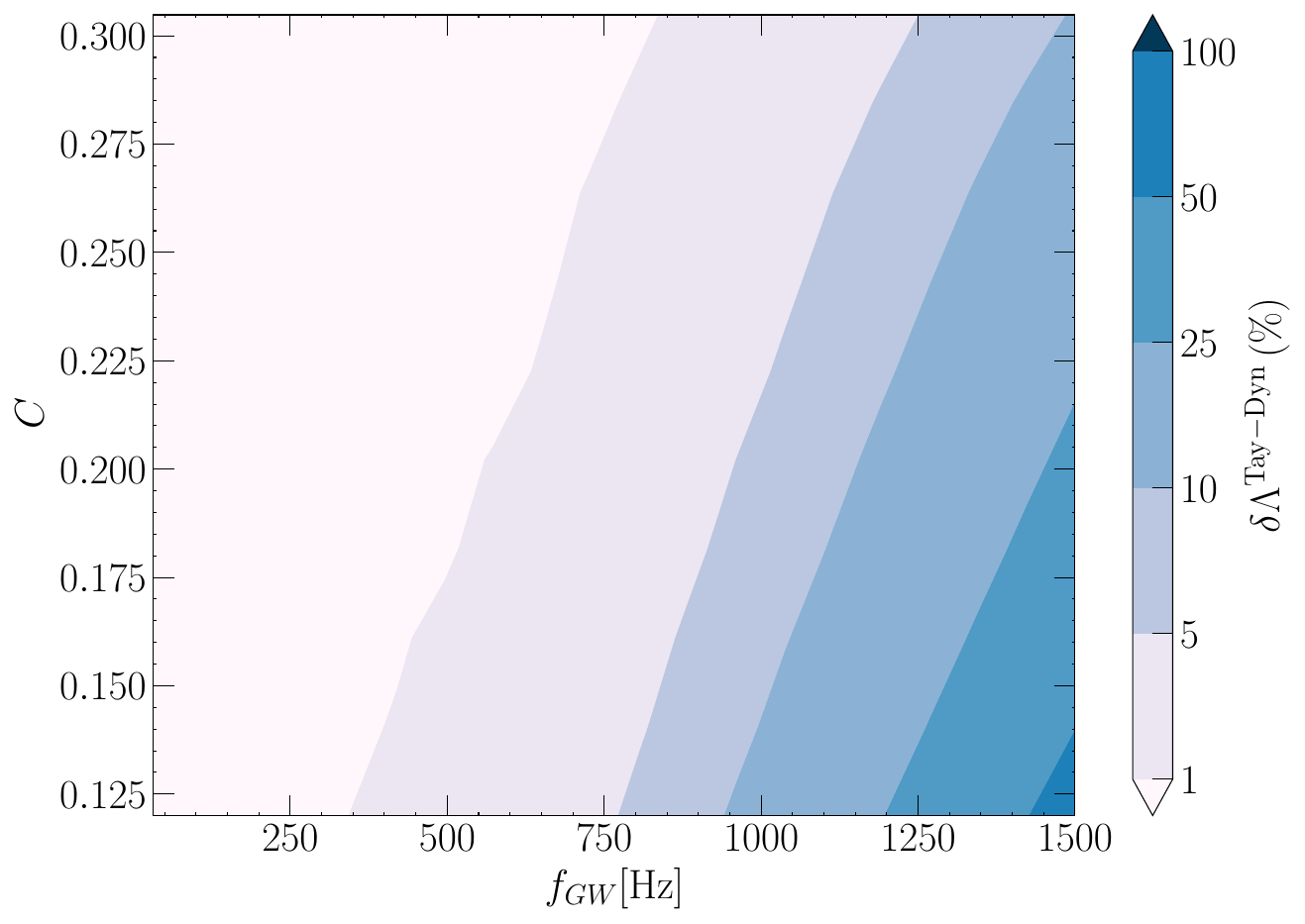}
    \quad
    \includegraphics[width=0.48\linewidth]{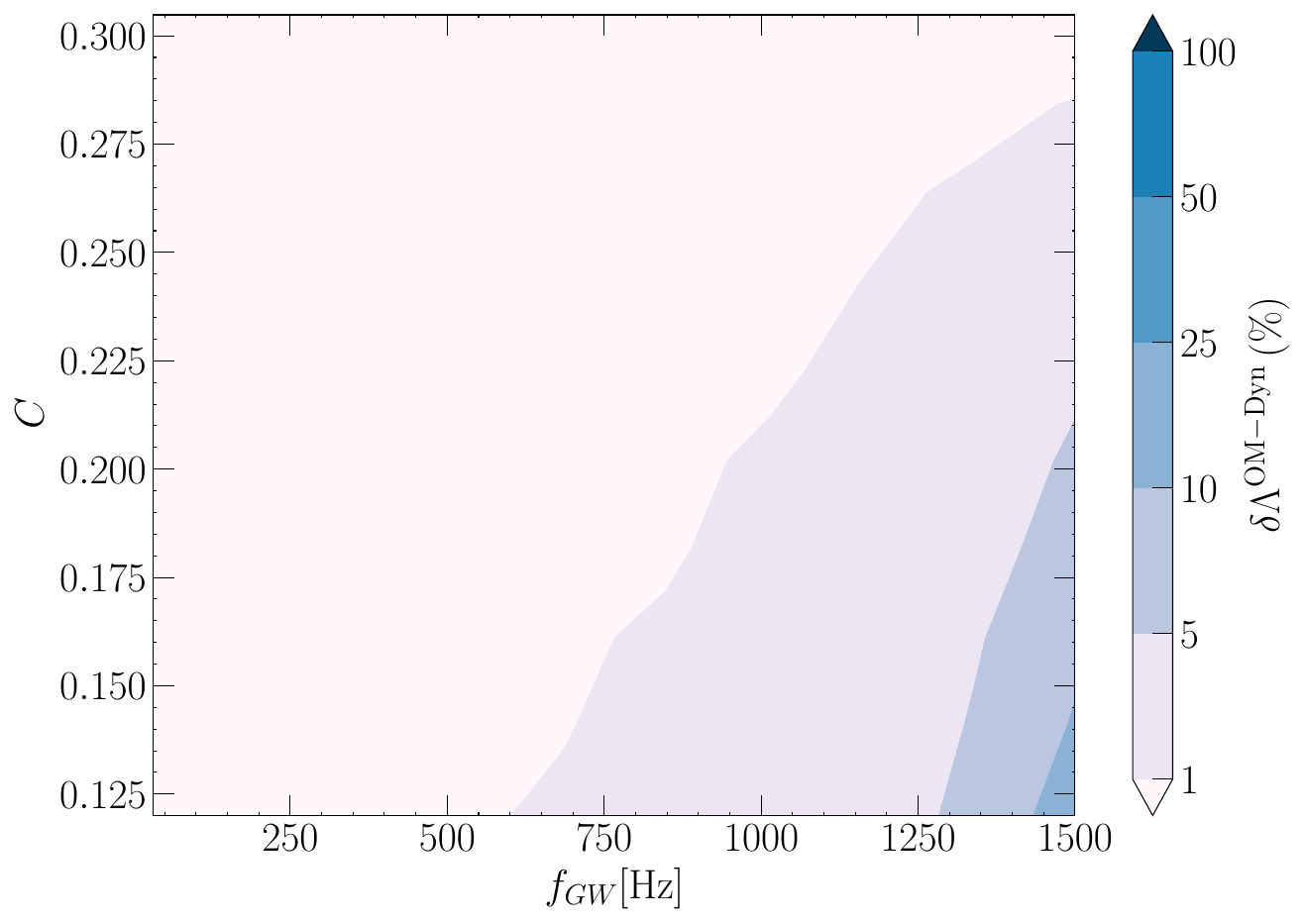}
    \caption{Percent fractional difference between the dynamical tide and the second-order small-frequency expansion (left) or the one-mode approximation (right), as a function of compactness and GW frequency. As the frequency increases, the Taylor series expansion gradually loses accuracy, with deviations remaining below $\sim 25\%$ across most of the parameter space. Only at high frequencies, combined with small compactness, does the error exceed this level, going up to $50\%$ in some regions. On the other hand, the one-mode approximation remains accurate at low frequencies, with errors below $5\%$ for $\omega \lesssim 1250\,\mathrm{Hz}$ and remaining below $25\%$ throughout the parameter space considered here.}
    \label{fig:lambdaOMlambdatrue}
\end{figure*}

We can better quantify the robustness of these approximations by calculating their relative difference to the dynamical tide across the $C$--$\omega$ plane. Let us define the relative fractional difference between the series expansion and the dynamical tide via
\begin{equation}
   \delta \Lambda^{\rm Tay-Dyn}= \Bigg | 1 -\frac{\Lambda^\text{Tay}(\omega)}{\Lambda^\text{Dyn}(\omega)}\Bigg |  = \Bigg | 1 -\frac{ \Lambda^{(0)}+ \Lambda^{(2)} (M \omega)^2}{\Lambda^\text{Dyn}(\omega)}\Bigg | ,
\end{equation}
which can then be evaluated across different frequencies and compactness for a given EOS. The results for the SLy EOS are shown in the left panel of Fig.~\ref{fig:lambdaOMlambdatrue}, from which we conclude that the series expansion is an accurate representation of the dynamical tidal response function over a large part of $C-f_{\mathrm{GW}}$ plane; the deviation remains below $\sim5\%$ for frequencies up to $f_{\text{GW}}\lesssim750$ Hz for the entire range of compactness considered. To obtain a relative error smaller than $10\%$ when using the Taylor dynamical tide for all compactness, one must limit the linear frequency to be below $\sim 1000$ Hz. This behavior is consistent with expectation, since the small-frequency expansion in $(M\omega)^2$ becomes less accurate as the frequency increases.

Let us now define the relative difference between the one-mode expansion and the dynamical tide as 
\begin{equation}
  \delta \Lambda^{\rm OM-Dyn}  = \Bigg | 1 -\frac{\Lambda^{\rm OM}(\omega)}{\Lambda^\text{Dyn}(\omega)}\Bigg |  =\Bigg |  1 -\frac{\Lambda^{(0)}(1-\omega^2/\omega_*^2)^{-1}}{\Lambda^\text{Dyn}(\omega)}\Bigg | ,
\end{equation}
which, again, we can evaluate across different frequencies and compactness for a given EOS. The results for the SLy EOS are shown in the right panel of Fig.~\ref{fig:lambdaOMlambdatrue}. From it we see that the one-mode approach demonstrates even higher agreement with the dynamical tide, with most of the error below $5\%$ for $f_{\rm GW}\lesssim1250$Hz. The disagreement between the two increases gradually for all compactness and reaches over 10\% for linear frequencies above 1400Hz.

These two results show us that both the Taylor series expansion in Eq.~\eqref{eq:approxdynamical} and the effective representation in Eq.~\eqref{eq:lambdaapproxonemode} provide a robust and quantitatively accurate approximation to the dynamical tidal deformability across a large regime of frequencies relevant to GW observations of binary NSs, with the OM approximation being more consistent with the dynamical approach for a larger range of frequencies. This suggests that all three are interesting approaches, which need to be further compared to numerical relativity simulations to better understand the systematic difference between them. Previous studies~\cite{Harry:2018hke} have shown that tidal effects begin to become noticeable in GW parameter estimation at linear frequencies larger than $400$ Hz. This is significantly earlier than the $f-$mode linear frequency of NSs, which, although dependent on the EOS, can be as low as $1.5$ kHz and potentially before contact. As the binary approaches these higher frequencies, the OM approximation of Eq.~\eqref{eq:lambdaapproxonemode} agrees better with the dynamical response than the Taylor description at all compactness.

\section{Universal Relations}
\label{sec:universalrelation}

Since the fundamental ($f-$)mode of oscillation of a NS is known to be universally related to the static tidal deformability~\cite{Chan:2014kua}, and because the numerical value of $\omega_*$ is similar to the $f-$mode value~\cite{Pitre:2023xsr}, one may wonder if the dynamical tidal deformability also presents some EOS-insensitive features. Let us then investigate the relation between the dimensionless static tidal deformability $\Lambda^{(0)}$ and its leading-order correction in frequency $\Lambda^{(2)}$, as well as the relation between $\Lambda^{(0)}$ and the dimensionless quantity $(M\omega_*)$ or, equivalently, $\sqrt{\Lambda^{(0)}/\Lambda^{(2)}}$. 

\begin{figure}[!ht]
    \centering
    \includegraphics[width=\linewidth]{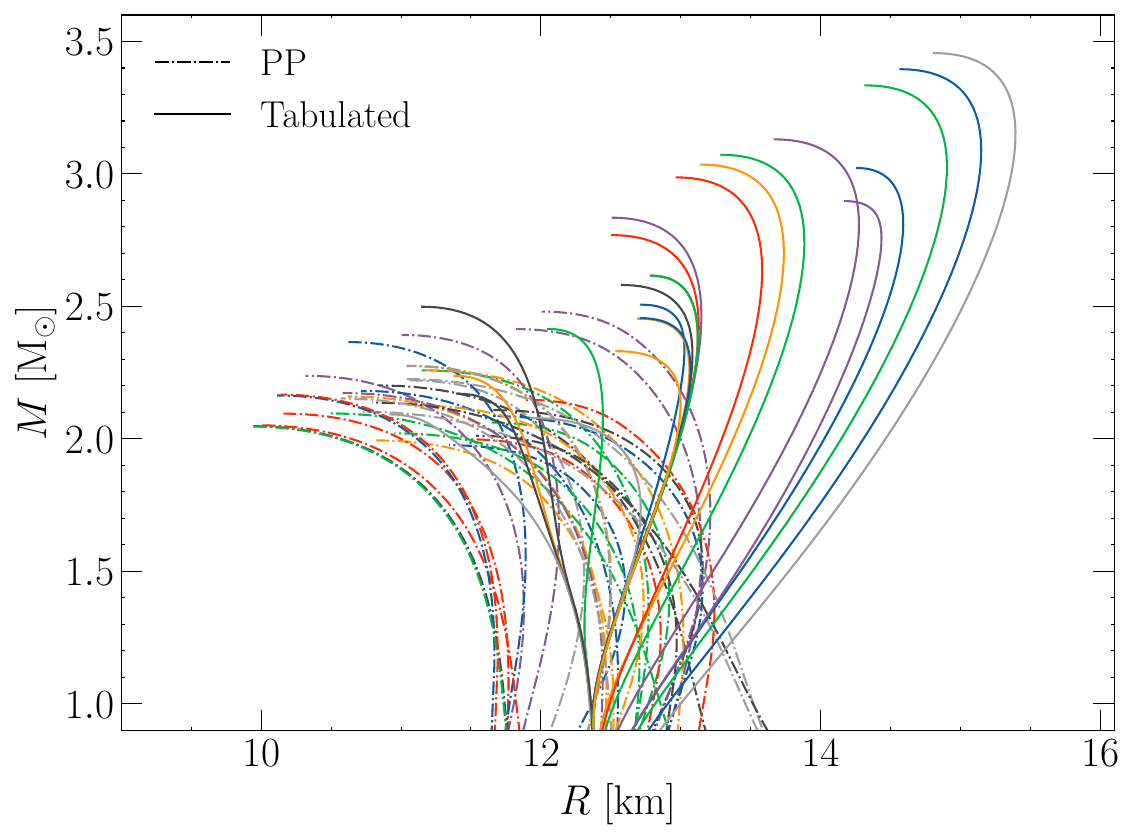}
    \caption{Mass-Radius curves for the set of EOS considered in this paper. Each curve corresponds to a sequence of NSs for a particular EOS. Along each curve, the central pressure increases until it reaches the maximum mass allowed for the particular EOS. Observe that the various EOSs lead to wide changes in the mass-radius curves, although they all pass observational constraints to 90\% confidence.}
    \label{fig:massradiuscurves}
\end{figure}

To assess the dependence of these relations on the EOS, we consider a broad sample of microphysical models and compute the corresponding $\Lambda^{(0)}$--$\Lambda^{(2)}$ and $\Lambda^{(0)}$--$M\omega_*$ curves for each case. The EOS set consists of the following two subsets: 
\begin{itemize}
    \item \textit{Physics-informed, piecewise polytropes (PP)}: 35 EOS models\footnotemark, parametrized as piecewise polytropes from~\cite{Read:2008iy};
    \item \textit{Phenomenological, tabulated EOSs}: 24 tables that exhibit non-smooth features in the speed of sound, specifically those studied in~\cite{Tan:2021ahl}.
\end{itemize}
\footnotetext[1]{Namely ALF2, AP3, BSK20–26, BSP, DD2, DD2Y, DDHd, DDME2, DDME2Y, ENG, FSUGarnet, G3, IOPB, MPA1, Model1, Rs, SINPA, SK255, SK272, SLY2, SLY230A, SLY4, SLY9, SLy, SkI4, SkI6, SkMP, Ska, Skb.}
The effect of this set on the mass-radius curve is shown in Fig.~\ref{fig:massradiuscurves}, in which each of the curves corresponds to a sequence of NSs with a different EOS. For each curve, the central pressure increases up to the maximum mass allowed by that EOS. All the EOSs selected here satisfy the 90\% constraints on the mass-radius plane set by the pulsars J0030+0451~\cite{Miller:2019cac} and J0740+6620~\cite{Miller:2021qha}, as well as the binary NS GW event GW170817~\cite{LIGOScientific:2018cki}. The set of EOSs studied here is by no means complete, but it serves as a representative sample that covers a wide range of observational possibilities.

\begin{figure}[!ht]
    \centering
    \includegraphics[width=\linewidth]{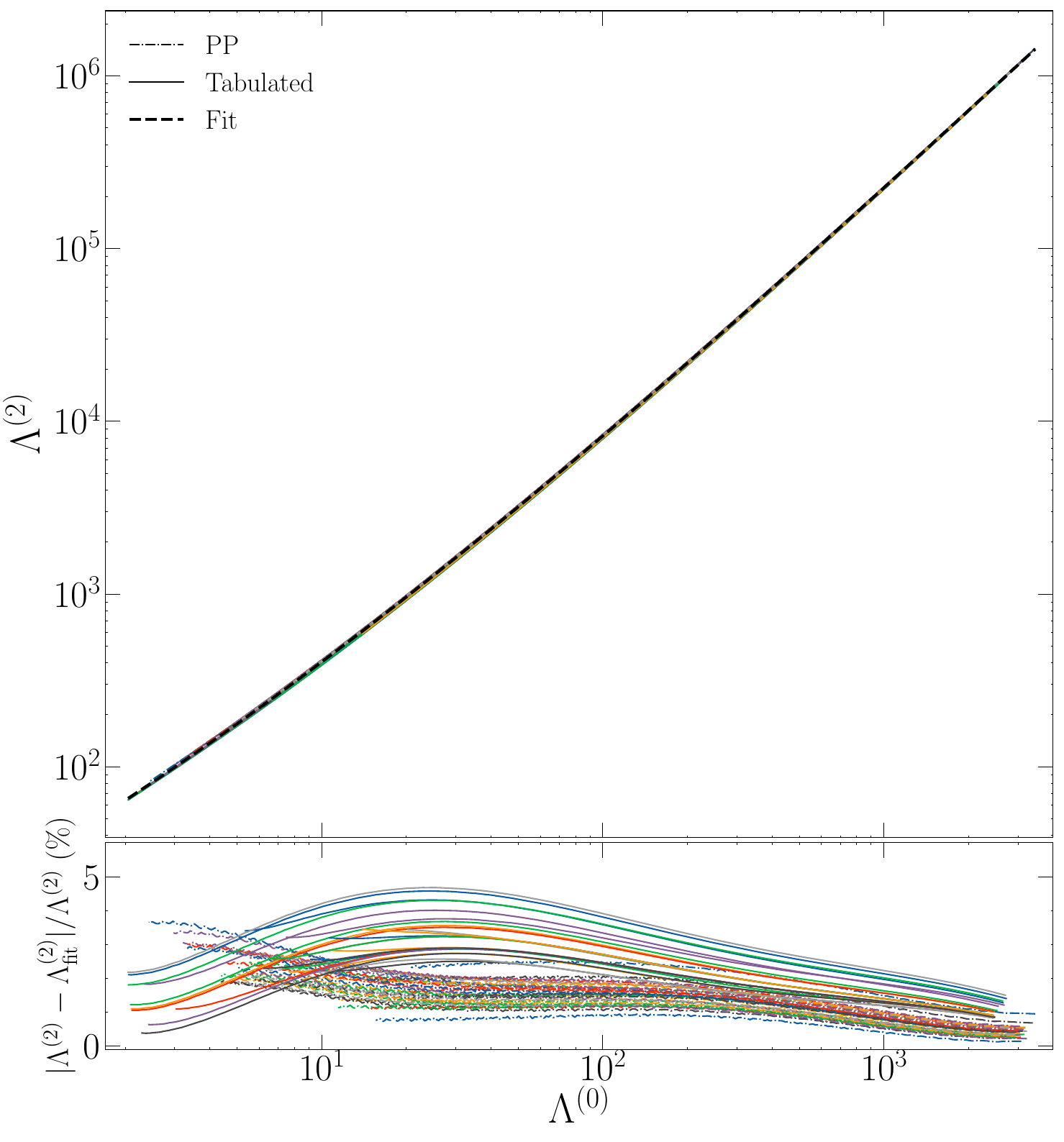}
    \caption{
    Relation between the dimensionless static tidal deformability $\Lambda^{(0)}$ and its second-order frequency correction $\Lambda^{(2)}$ for the complete set of EOS. Compactness varies along each curve from the maximum allowed mass (left corner) down to $C=0.12$ (right corner).  
    \textbf{Top:} Relation $\Lambda^{(2)}$--$\Lambda^{(0)}$  in log-log scale. Here, solid lines correspond to the tabulated EOS, dash-dotted lines to the PP EOS set, and the thick dashed line to the fit relation~\eqref{eq:lambda_fit}. 
    \textbf{Bottom:} Fractional deviation from the best fit, expressed in percent. Observe that the maximum relative difference to the fit is $\sim 5\%$.
    }
    \label{fig:lambda_relation}
\end{figure}

The $\Lambda^{(2)}$--$\Lambda^{(0)}$ relation computed for the complete set of EOSs is shown in Fig.~\ref{fig:lambda_relation}. Each curve corresponds to a single EOS, with compactness varying along the curve from the maximum supported mass down to $C=0.12$. Despite the diversity in the microphysics and in the phenomenology, observe that the curves all lie within a narrow band, indicating a quasi-universal relation between $\Lambda^{(2)}$ and $\Lambda^{(0)}$ across the considered EOSs.

We quantify this universality by fitting the combined dataset to the functional form
\begin{equation}
    \ln \Lambda^{(2)} = a_0 + a_1 (\ln \Lambda^{(0)})+ a_2 (\ln \Lambda^{(0)})^2 + a_3 (\ln \Lambda^{(0)})^ 3,
    \label{eq:lambda_fit}
\end{equation}
where $\ln$ stands for the natural logarithm, $a_0$, $a_1$, $a_2$ and $a_3$ are free parameters determined via least-squares regression. The best-fit parameters for the complete EOS set are $ (a_0,a_1,a_2,a_3) = (3.446, 0.9876, 5.811\times 10^{-2}, -2.174\times 10^{-3})$ and the fractional differences between the true values and the fit values are shown in Fig.~\ref{fig:lambda_relation}. Observe that the $\Lambda^{(2)}$--$\Lambda^{(0)}$ relation is indeed near EOS-independent, with $90\%$ of the ``error'' around $3\%$, and the maximum error at around $5\%$. This level of quasi-universality is comparable to that of the original I-Love-Q relations~\cite{Yagi:2013awa}, and much greater than that of the binary Love relations~\cite{Yagi:2015pkc} (by a factor of $3$--$6$). 

This result suggests that $\Lambda^{(2)}$ can be calculated directly from $\Lambda^{(0)}$ with minimal dependence on the underlying EOS. Consequently, utilizing the relation in Eq.~\eqref{eq:lambda_fit} avoids the need to introduce additional free parameters in parameter estimation to infer $\Lambda^{(2)}$. That is, we can use the relation in Eq.~\eqref{eq:lambda_fit} to write the dynamical tidal response entirely in terms of $\Lambda^{(0)}$ when estimating parameters from GW observations; then, once the inference is done, we can use Eq.~\eqref{eq:lambda_fit} again to infer \textit{a posteriori} the value of $\Lambda^{(2)}$, and thus, make further inferences on nuclear physics (related to the microphysics that impact $\Lambda^{(2)}$) that would not be possible otherwise. 

\begin{figure}[!ht]
    \centering
    \includegraphics[width=\linewidth]{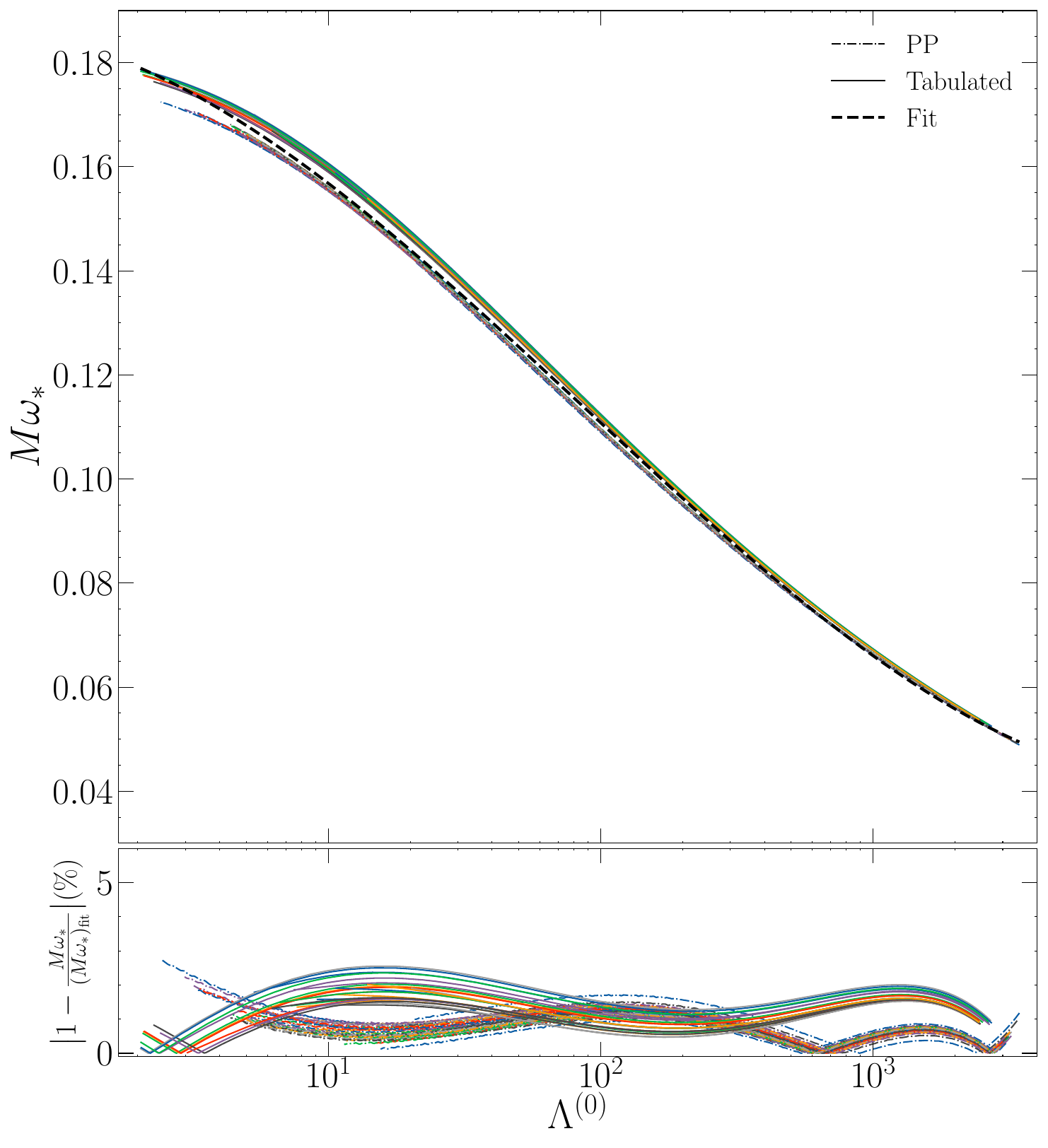}
    \caption{
    Relation between the dimensionless static tidal deformability $\Lambda^{(0)}$ and the effective dimensionless frequency $M \omega_*$ for the complete set of EOSs. Compactness varies along each curve from the maximum allowed mass (left corner) down to $C=0.12$ (right corner).  
    \textbf{Top:} Relation $\Lambda^{(0)}$--$M \omega_*$. Here, solid lines correspond to the tabulated EOS, dash-dotted lines to the PP set EOS, and the thick dashed line to the fit relation~\eqref{eq:omega_fit}. 
    \textbf{Bottom:} Fractional deviation from the best fit, expressed in percent. Observe that the maximum deviation from the fit is $\sim 2.8\%$.
    }

    \label{fig:lambda_y_relation}
\end{figure}

The universality of these two parameters can also be studied through the relation between $\Lambda^{(0)}$ and $M \omega_*$. Using the values of $\Lambda^{(0)}$ and $\Lambda^{(2)}$ computed from each EOS, we can calculate $M \omega_*$ through Eq.~\eqref{eq:omega*}, and then compare this to $\Lambda^{(0)}$. The $\Lambda^{(0)}$--$M \omega_*$ relation is presented in Fig.~\ref{fig:lambda_y_relation}, which again reveals a strong quasi-universality. We quantify the EOS-dependency by fitting the combined dataset to the functional form
\begin{equation}
    M \omega_* = b_0 + b_1 (\ln \Lambda^{(0)})+ b_2 (\ln \Lambda^{(0)})^2 + b_3 (\ln \Lambda^{(0)})^ 3,
    \label{eq:omega_fit}
\end{equation}
where $b_0$, $b_1$, $b_2$ and $b_3$ are free parameters determined via least-squares regression. The best-fit parameters for the complete set of EOSs are $(b_0, b_1, b_2, b_3) = (0.1843, -5.025 \times 10^{-3}, -3.621\times 10^{-3}, 2.708\times 10^{-4})$, and the fractional error between the true values and the fit values is shown in Fig.~\ref{fig:lambda_y_relation}. The figure shows that the $\Lambda^{(0)}$--$M \omega_*$ relation is indeed near EOS-independent, with the $90\%$ of error smaller than $2\%$, and the maximum error around $2.8\%$. 

Throughout this paper we calculate the values of $\Lambda^{(2)}$ using the normalization for the external solutions which is presented in~\cite{HegadeKR:2024agt} and described in Appendix~\ref{Ap:BCs}. For completeness, we also present the universal relations between $\Lambda^{(2)}$--$\Lambda^{(0)}$ and $\Lambda^{(0)}$--$M\omega_*$ calculated using the external solutions introduced in~\cite{Pitre:2023xsr}. The plots alongside the best-fit parameters are presented in~\ref{Ap:Normalization}.

\section{Conclusion}

In this work, we presented a detailed method to compute $\Lambda^{(2)}$, the leading-order correction in frequency to the static tidal deformability, following the pioneering works of~\cite{Pitre:2023xsr,HegadeKR:2024agt}. We quantified the contribution of this frequency correction by first comparing it to the static tide. Our findings show that the Taylor-expanded approach proposed here improves substantially over the static tide. Moreover, we compared the dynamical tide to both the Taylor-expanded approach and a one-mode approach and found that the former can be accurately captured by both approximations, with relative differences below $10$\% for $f_{\text{GW}}\lesssim 1000$ Hz. At higher frequencies, the one-mode approach has better agreement with the dynamical tide than the Taylor approach for all compactnesses.

We further identified and quantified two novel quasi-universal relations between the static tidal deformability $\Lambda^{(0)}$ and its leading-order dynamical correction $\Lambda^{(2)}$, as well as between the former and the effective angular frequency $M\omega_*$. We performed our analysis over a set of 59 EOSs, which include both physics-informed, piecewise polytropic representations~\cite{Read:2008iy} and phenomenological, tabulated EOSs that contain non-trivial behavior in the speed of sound~\cite{Tan:2021ahl}, some with maximum masses of $3$ M$_{\odot}$. The $\Lambda^{(0)}$--$\Lambda^{(2)}$ relation presents strong universality, with EOS variability of at most $\sim 5\%$ and with 90\% of the non-universality below $3\%$. The $\Lambda^{(0)}$--$M\omega_*$ relation reveals an even smaller EOS sensitivity, holding universally to within $\sim 2.8\%$ and having 90\% of the non-universality below $2\%$. Such a high degree of EOS insensitivity is present only in a select few quasi-universal relations, such as the I-Love-Q relations~\cite{Yagi:2013awa} and the $f-$mode--$\Lambda^{(0)}$ relations~\cite{Chan:2014kua}.

It is important to note that the approximations of the dynamical tide and the quasi-universal relations presented here come with several caveats. In particular, this framework does not capture resonances associated with low-frequency $g$-modes~\cite{Lai:1993di,Yu_2016,Kuan:2021jmk,Kuan:2021sin,Counsell:2024pua,Reboul-Salze:2025gyi} or interface ($i-$)mode excitations~\cite{Counsell:2025hcv,Gao:2025aqo}. These resonances may arise within the frequency band relevant to binary inspiral and could modify the tidal response in a non-negligible way. At higher frequencies, resonant excitation of the $f$-mode may occur prior to merger; such effects lie outside the scope of the present analysis and are not represented by either parametrization considered here. More broadly, our framework neglects additional physical ingredients that could introduce further structure into the tidal response, such as spin effects~\cite{Kuan:2022etu,Kuan:2022etu,Yu:2024uxt}, which could also be studied in the future.

The relations presented here establish a first, direct, and universal connection between the static and dynamical tidal responses of NSs, but more work can always be done. Effects caused by a first-order phase transition, for example, which can also generate $i-$modes~\cite{Counsell:2025hcv}, are not incorporated into the set of EOSs considered in this work and may introduce a deterioration of the universality of the relations presented here. Future work could focus on enlarging the EOS space and quantifying and modeling carefully any remaining EOS sensitivity.  

By expressing the leading-order dynamical correction $\Lambda^{(2)}$ in terms of the adiabatic tidal deformability ${\Lambda^{(0)}}$, we presented a method for capturing features of the dynamical tides without introducing additional parameters in data analysis. This approach, similar to the one employed when using the binary Love Relation~\cite{Yagi:2016qmr}, offers a route to incorporate the frequency-dependent tidal effects into GW models without the need for extra parameters. Further work could focus on understanding the degree to which parameter estimation of $\Lambda^{(0)}$ is enhanced when using the universal relations presented here. In a practical implementation, one will probably have to fold in the EOS variability of the quasi-universal relations through a Gaussian representation of the error that one then marginalizes over, as done currently when employing the binary Love relations~\cite{Chatziioannou:2018vzf}.

The universal relations presented here could also be used as a probe of modified gravity in a way that is EOS-insensitive. Modified gravity theories are likely to lead to EOS-insensitive relations, but these may differ from those presented here; this ought to be verified in practice by studying specific modified theories that have not yet been ruled out, like dynamical Chern-Simons gravity~\cite{Jackiw:2003pm,Alexander:2009tp}. The distance between the modified-gravity EOS-insensitive relations and the general-relativity EOS-insensitive relations would then be proportional to the coupling constants of the modified theory. A combined measurement of two parameters in these universal relations (such as $\Lambda^{(0)}$ and $\omega_*$) would then enable an interesting test of general relativity that is approximately independent of the unknown microphysics at work inside NSs. This scenario is precisely what happens with the I-Love-Q relations~\cite{Yagi:2013awa}, which have already led to stringent tests of general relativity~\cite{Silva:2020acr}. An independent measurement of $\Lambda^{(0)}$ and $\omega_*$ will, of course, require a large signal-to-noise ratio event, and determining the latter would also constitute an interesting line for future work.

\section{Acknowledgements}\label{sec:acknowledgements}
The authors acknowledge support from the Simons
Foundation through Award No.~896696, the Simons Foundation International through Award No.~SFI-MPS-BH-00012593-01, the NSF through Grants No.~PHY-2207650 and~PHY-25-12423, and NASA through Grant No. 80NSSC22K0806.

The authors would also like to thank Hang Yu for the valuable and insightful comments on this manuscript.

\bibliographystyle{apsrev4-2} 
\bibliography{lib}

\appendix

\section{Coefficients of Small Frequency Master Equations}
\label{Ap:Coefs}

Here, we explicitly present the coefficients of the small-frequency master equations. These expressions depend on the radial coordinate $r$, the equilibrium fluid variables $p(r)$ and $\epsilon(r)$, and the background metric functions $\lambda(r)$ and $\nu(r)$, defined by the line element: 
\begin{align}
\label{eq:lineelement}
ds^2 = -e^{\nu(r)}\,dt^2 +e^{\lambda(r)} dr^2 
 +\, r^2\, (d\theta^2 + \sin{\theta}^2 d\phi^2).
\end{align}
The coefficients from Eq.~\eqref{eq:masterexp} are given below:
\begin{align}
    &\alpha _{H,0}^{(0)}= \frac{1}{r^2}\Bigg\{\left(8 \pi  r^2 p+1\right)^2 e^{2 \lambda }-3+ \nonumber\\ &e^{\lambda } \left[ 2-4 \pi  r^2 \left(\frac{d\epsilon}{dp}(p+\epsilon)+3\epsilon+15p\right)\right] \Bigg\}
    \nonumber \\
    &\alpha _{H,1}^{(0)}= \alpha _{H,2}^{(0)} = 0\\
    &\alpha _{H,3}^{(0)}=\frac{e^{\lambda } \left[4 \pi  r^2 \epsilon-4 \pi  r^2 p-1\right]-5}{r}\nonumber 
\end{align}

\begin{align}
    \alpha _{W,0}^{(0)}&= -\frac{r \left\{\left(8 \pi  r^2 p+1\right)^2 e^{2\lambda }+2 e^{\lambda } \left[4-4 \pi  r^2 (\epsilon+ p)+ \frac{d\epsilon}{dp}\right]-3\right\}}{2 e^{\lambda/2}}
    \nonumber \\
    \alpha _{W,1}^{(0)}&=\frac{\left[\left(8 \pi  r^2 p+1\right) e^{\lambda }-1\right]  \frac{d\epsilon}{dp}-6}{2 r} \\
    \alpha _{W,2}^{(0)}&= -\frac{6 e^{\lambda/2}}{r}\nonumber\\
    \alpha _{W,3}^{(0)}&=-\frac{r^2 \left[\left(8 \pi  r^2 p+1\right) e^{\lambda }-1\right]}{2 e^{\lambda/2}}\nonumber 
\end{align}

\begin{align}
    \alpha _{V,0}^{(0)}&= \frac{r}{12 e^{\lambda }} \Bigg\{\left(8 \pi  r^2 p+1\right)^3 e^{3\lambda }\nonumber\\&-\left(8 \pi  r^2 p+1\right) e^{2\lambda } \left[8 \pi  r^2 \left(\epsilon+4p\right) +5\right]\nonumber\\&+3 e^{\lambda } \left(8 \pi  r^2 \epsilon-7\right)+9\Bigg\}
    \nonumber \\
    \alpha _{V,1}^{(0)}&=\frac{e^{\lambda/2} \left(8 \pi  r^2 \epsilon+8 \pi  r^2 p-3\right)}{3 r}\\
    \alpha _{V,2}^{(0)}&=\frac{\left(8 \pi  r^2 p+1\right) e^{\lambda }-3}{r}\nonumber\\
    \alpha _{V,3}^{(0)}&=\frac{r^2 \left[-\left(8 \pi  r^2 p+1\right)^2 e^{2\lambda }-4 \left(8 \pi  r^2 p+3\right) e^{\lambda }+3\right]}{12 e^{\lambda }}\nonumber 
\end{align}

\begin{align}
  \alpha _{H,0}^{(2)}&= \frac{e^{-\nu }}{12 e^{\lambda }} \Bigg\{  8 \left(8 \pi  r^2 p+1\right)^2 e^{3\lambda } \left(\pi  r^2 \epsilon+7 \pi  r^2 p+1\right)- \nonumber\\&e^{4\lambda }\left(8 \pi  r^2 p+1\right)^4 \nonumber\\&-72 e^{\lambda } \left(-\pi  r^2 \epsilon+\pi  r^2 p+1\right)
   \nonumber\\&-6 e^{2\lambda }\left[ 8 \pi  r^2 \left(8 \pi r^2 p (\epsilon+2 p)+\epsilon+ p\right)+1\right]+27\Bigg\}
    \nonumber \\
   \alpha _{H,1}^{(2)}&=\frac{-8\pi  e^{\lambda/2-\nu} }{3}   \left[\left(8 \pi  r^2 p+1\right) e^{\lambda }-3\right] (\epsilon+p)\\
   \alpha _{H,2}^{(2)}&=4 \pi  e^{\lambda-\nu } (\epsilon +p) \left( \frac{d\epsilon}{dp}-1\right)\nonumber\\
   \alpha _{H,3}^{(2)}&=-\frac{r e^{-\lambda-\nu }}{12}\left[\left(8 \pi  r^2 p+1\right) e^{\lambda }-3\right] \nonumber \\ & \Bigg\{ \left(8 \pi  r^2 p+1\right)^2 e^{2\lambda }-4 \left(8 \pi  r^2 p+3\right) e^{\lambda }+3\Bigg\}.\nonumber 
\end{align}

\section{Boundary Conditions \& Exterior Solutions}
\label{Ap:BCs}

This appendix contains the boundary conditions necessary to integrate the master equations [see Eq.~\eqref{eq:masterexp}]. We start by describing the behavior of the functions near the center of the star; then, we discuss the solutions near the surface of the star, and lastly, we present the exterior solutions. 

\subsection{Near the center}
\label{sec:Nearthecenter}

Due to the singular behavior of the equations near the origin, direct numerical integration from $r = 0$ is not feasible. Instead, the integration is initiated at a small but finite radial coordinate $r = r_0 $.  To provide consistent initial conditions at $r = r_0 $, we perform a power series expansion of the relevant functions about the origin. This expansion captures the leading-order behavior of the variables under the requirement of regularity at the center. The resulting expressions are then evaluated at $r = r_0$, supplying the necessary starting values for the numerical integration of the system. Since the perturbation functions  $H^{(0)}(r)$, $W^{(0)}(r)$, $V^{(0)}(r)$, and $H^{(2)}(r)$ are sourced by the background configuration of the star, we also require regularity of the background fields at the origin. To enforce this, we begin by analyzing the behavior of the equilibrium quantities described by the TOV equations. 

\subsubsection{Background solutions}

The equilibrium configuration of the NS is described by several functions, such as the pressure $p$, the enclosed mass $m$, the metric functions $\nu$ and $\lambda$, and the energy density $\epsilon$. Regularity of the background functions near the origin requires that all physical and metric quantities admit smooth power series expansions in $r$ when $r\approx 0$. Substituting these into the TOV equations yields
\begin{align}
    m(r) &= m_3 r^3 + \mathcal{O}(r^4), \nonumber \\
    p(r) &= p_0 + p_2 r^2 + \mathcal{O}(r^3), \nonumber \\
    \epsilon(r) &= \epsilon_0 + \epsilon_2 r^2 + \mathcal{O}(r^3),  \\
    \nu(r) &= \nu_0 + \nu_2 r^2 + \mathcal{O}(r^3), \nonumber \\
    \lambda(r) &= \lambda_2 r^2 + \mathcal{O}(r^3),\nonumber 
\end{align}
where the coefficients $m_3$, $p_2$, $ \epsilon_2$, $\nu_2$, and $\lambda_2$ are determined by expanding the TOV equations about $r=0$, they are given by: 
\begin{align}
    m_3&=  \frac{4}{3} \pi \epsilon_0, \nonumber \\
    p_2 &= -\frac{2\pi}{3} (\epsilon_0+p_0) (\epsilon_0 + 3 p_0), \nonumber \\
    \epsilon_2 &= p_2  \frac{d\epsilon}{dp}\Big|_{r=0},  \\
    \nu_2 &= \frac{4\pi}{3} (\epsilon_0 + 3 p_0), \nonumber \\
    \lambda_2 &= 2m_3. \nonumber 
\end{align}
The central pressure $p_0$ serves as a free parameter, while $\epsilon_0$ is fixed by the chosen equation of state and $\nu_0$ is determined by matching $\nu(r)$ at the surface with the Schwarzschild metric.

\subsubsection{Perturbation equations}

For the leading-order metric perturbation function $H^{(0)}(r)$, we obtain that
\begin{equation}
\label{eq:smallrH}
H^{(0)}(r) = h^{(0)}_0 \left(1 + \mathcal{O}(r^2)\right), 
\end{equation} 
where $h^{(0)}_0$ is a scaling factor that appears due to the homogeneity of the equation. Similarly, for the fluid perturbation variables $ W^{(0)}(r)$ and $V^{(0)}(r)$, we write a series solution of the form 
\begin{equation}
W (r)= w_0 + w_2 r^2 + {\cal{O}}(r^4), \quad
V (r) = v_0 + v_2 r^2 + {\cal{O}}(r^4),
\end{equation}
and regularity at the center implies that the coefficients admit the following expressions 
\begin{align}
\label{eq:smallrVW}
w_0 &= -2 v_0, \nonumber \\ 
w_2 &= -  \frac{2}{7 p_2} h_{0}^{(0)} ( p_2+\epsilon_2) \nonumber \\
&-  \frac{2}{7 p_2} v_{0} \left[4 \pi  p_0 ( p_2+2 \epsilon_2)+6 \lambda_2 p_2-8 \pi  p_2 \epsilon_0+\lambda_2 \epsilon_2\right],\nonumber \\
v_2 &=  \frac{1}{42 p_2} h_{0}^{(0)} (3 \epsilon_2-11 p_2) \nonumber \\ 
& +  \frac{1}{42 p_2}v_{0} \left[8 \pi  p_0 (5 p_2+3 \epsilon_2)+39 \lambda_2 p_2-80 \pi  p_2 \epsilon_0+3 \lambda_2 \epsilon_2\right],\nonumber 
\end{align}
where $v_0$ is a free constant. For the dynamical correction to the metric function $H^{(2)}(r)$, we have again an ansatz of the form
\begin{equation}
    H^{(2)}(r)= h^{(2)} _0 + h^{(2)} _2 r^2 + {\cal{O}}(r^4).
\end{equation} 
Using this in the perturbed equations, we find that 
\begin{align}
     h^{(2)} _2 =&\frac{1}{42} e^{-\nu_{0}} (-11 h^{(0)}_0-12 h^{(2)}_0 e^{\nu_{0}} (-\lambda_2+2 \pi  p_0^3 \epsilon_2^2+5 \pi  p_0^2 \epsilon_2+\nonumber \\ &\pi  p_0 (2 \epsilon_{0} \epsilon_2+11)+3 \pi  \epsilon_{0})+\nonumber \\
     &4 \pi  \left(p_0^2 \epsilon_2+p_0+\epsilon_{0}\right) (v_0 (6 p_0 \epsilon_2-3)+4 w_0)),
\end{align}
where we again have a free constant $h^{(2)}_0$. 

\subsection{Near the surface}

As the radial coordinate approaches the stellar radius $r = R$, the pressure $p(r)$ drops to zero, defining the boundary of the star. Near this surface, the structure of the background solution remains smooth, but care must be taken in handling the vanishing pressure, particularly when evaluating terms that involve $p(r)$ or the speed of sound $c_s^2 = dp/d\epsilon$ in the denominator. For example, due to the vanishing of the speed of sound at the surface, certain perturbation equations become numerically unstable (if integrated outward from the center up to the surface). To address this, some functions are instead integrated inward from the surface to an intermediate matching point $r=r^*$, and also integrated outward from the center to the same matching radius. These two solutions are then smoothly joined at this midpoint to ensure continuity.

Although the equation governing the metric perturbation $H^{(0)}(r) $ remains well-behaved at the stellar surface (and can thus be integrated directly from the center to the surface without issue), the fluid perturbation functions require special treatment to ensure regularity at $ r = R$. In particular, to avoid divergences in terms involving $1/c_s^2$, one must impose the condition that the Lagrangian variation of the pressure vanishes at the surface, which provides the appropriate boundary condition for the fluid sector.
This leads to
\begin{align}
    V^{(0)}(r) =& v_R + {\cal{O}}[(r-R)^2] \nonumber \\
    W^{(0)}(r) =& \frac{2 \,R^2 \,H^{(0)}(R) \,e^{\lambda(R)/2}}{8\, \pi \, R^2 \,p(R) \, e^{\lambda (R)}+e^{\lambda (R)}-1} + {\cal{O}}[(r-R)^2] 
\end{align}
where $ v_R $ is a free parameter. The expression for $ W^{(0)}(R) $ ensures that singular terms are canceled out and maintains the regularity of the perturbation equations at the boundary. 
The dynamical correction $ H^{(2)}(r) $ can in principle be integrated from the center to the surface; however, due to numerical instabilities, we adopt the same matching strategy at $ r = r^*$. The surface condition for this integration is given by 
\begin{align}
    H^{(2)}(r=R) = h^{(2)}_R + {\cal{O}}[(r-R)^2]\,,   \\
    H^{(2)'}(r=R) = h^{(2)'}_R + {\cal{O}}[(r-R)^2] \,,
\end{align}
where both constants are free parameters.

\subsection{Exterior Solutions}

The exterior solutions for the perturbation functions are derived by solving the master equations outside the star. Since the fluid perturbation functions go to zero outside the star, we don't consider their behavior in the exterior region.
 The methodology we employ to obtain the exterior behavior of our specific choice of metric perturbation functions, $H^{(0)}(r)$ and $H^{(2)}(r)$, is detailed in~\cite{HegadeKR:2024agt}. Here, we present the final expressions in our notation. 

We begin by considering the general solution for $H(r,\omega)$ up to second order in $\omega$, namely 
\begin{equation}
\label{eq:exteriorHfull}
    H(r,\omega) = 
    \frac{4\pi}{5}
    d_{2m}(\omega)
    \left[
    k_2(\omega) \mathbb{H}_{\rm multipole}+ \mathbb{H}_{\rm tide}
    \right],
\end{equation}
where $d_{2m}$ is the tidal moment of the source, $\mathbb{H}_{\rm tide}$ is associated with the tidal field, and $\mathbb{H}_{\rm multipole}$ corresponds to the part of the functions associated with the tidal response of the star. The explicit expressions of both of these functions are: 
\begin{widetext}
\begin{align}\label{eq:Htide-ours}
    \mathbb{H}_{\rm tide} &= z+ 
    \frac{M^2 \omega ^2 }{20160 (z-1)^2\, z}\Bigg\{80640 (z-1)^2\, z^2 \left[-\text{Li}_2(z)-\log (1-z) \log (z)+\frac{\pi ^2}{6}\right]-3029 \,z^6-6720 \,z^6 \log (z)+37010\, z^5\\
    &+67200 \,z^5 \log (z) +10555\, z^4-40320 z^4 \log ^2(z)-191676 \,z^4 \log (z)-
    208448\, z^3+80640 z^3 \log ^2(z)+\nonumber\\
    &154872 \,z^3 \log (z)+308389\,z^2-40320\, z^2 \log ^2(z)+36804 \,z^2 \log (z)+192 (z-1)^2 \log \left(\frac{1-z}{2}\right)\nonumber\\
    & \left(35 \,z^4-280 \,z^3+214\, z^2+420 \,z^2 \log (z)+280 \,z-35\right)-174290\, z-67200 \,z \log (z)+6720 \log (z)+8693\Bigg\}\nonumber
    \,,
\end{align}

\begin{align}\label{eq:Hmulti-ours}
    \mathbb{H}_{\rm multipole} &=\frac{-5 \left(z^4-8 z^3+12 z^2 \log (z)+8 z-1\right)}{32 C^5 z} + \frac{M^2 \omega ^2 }{70560 C^5 (z-1)^2 z}\Bigg\{-1058400 z^4 \text{Li}_3(z)+2116800 z^3 \text{Li}_3(z)\\
    &-1058400 z^2 \text{Li}_3(z)+1641780 (z-1)^2 z^2 \left(-\text{Li}_2(z)-\log (1-z) \log (z)+\frac{\pi ^2}{6}\right)-44100 (z-1)^2 \text{Li}_2(z)\nonumber \\
    &\left(z^4-8 z^3-25 z^2-12 z^2 \log (z)+8 z-1\right)+7350 \pi ^2 z^6+54233 z^6+22050 z^6 \log ^2(z)-22470 z^6 \log (2)\nonumber\\
    &+22470 z^6 \log (1-z)-44100 z^6 \log (1-z) \log (z)-22470 z^6 \log (z)-73500 \pi ^2 z^5-563960 z^5\nonumber\\
    &-220500 z^5 \log ^2(z)+224700 z^5 \log (2)-224700 z^5 \log (1-z)+441000 z^5 \log (1-z) \log (z)+180600 z^5 \log (z)\nonumber\\
    &+1058400 z^4 \zeta (3)-58800 \pi ^2 z^4+761171 z^4+88200 z^4 \log ^3(z)+374850 z^4 \log ^2(z)-381990 z^4 \log (2)\nonumber\\
    &+381990 z^4 \log (1-z)+88200 \pi ^2 z^4 \log (z)-269640 z^4 \log (2) \log (z)+622440 z^4 \log (1-z) \log (z)\nonumber\\
    &+687756 z^4 \log (z)-2116800 z^3 \zeta (3)+367500 \pi ^2 z^3+906500 z^3-176400 z^3 \log ^3(z)-176400 \pi ^2 z^3 \log (z)\nonumber\\
    &+539280 z^3 \log (2) \log (z)-2744280 z^3 \log (1-z) \log (z)-1585932 z^3 \log (z)+1058400 z^2 \zeta (3)-308700 \pi ^2 z^2\nonumber\\
    &-2218921 z^2+88200 z^2 \log ^3(z)-374850 z^2 \log ^2(z)+381990 z^2 \log (2)-381990 z^2 \log (1-z)+88200 \pi ^2 z^2 \log (z)\nonumber\\
    &-269640 z^2 \log (2) \log (z)+2121840 z^2 \log (1-z) \log (z)+173046 z^2 \log (z)+73500 \pi ^2 z+1029460 z\nonumber\\ &+220500 z \log ^2(z)-22050 \log ^2(z)-224700 z \log (2)+224700 z \log (1-z)-441000 z \log (1-z) \log (z)\nonumber\\ &+720300 z \log (z)-22470 \log (1-z)+44100 \log (1-z) \log (z)-14700 \log (z)-7350 \pi ^2+31517+22470 \log (2)\Bigg\}\nonumber
    \,,
\end{align}
\end{widetext}
where $z = 1 - 2M/r$.

\section{Matching}
\label{Ap:Matching}

To obtain a solution for the perturbation variables that is valid in the entire interior of the star, we perform a matching procedure at two locations: an interior matching point $ r = r^*$, and the surface of the star $r = R$. This allows us to stabilize the numerical integration and determine the free constants introduced in the near-center and near-surface expansions. In this subsection of this appendix, we separately address the matching procedures for each integration: the static metric perturbation, the fluid perturbations, and the dynamical correction to the metric perturbation.
\begin{figure}[h!]
    \centering
    \includegraphics[width=0.95 \columnwidth]{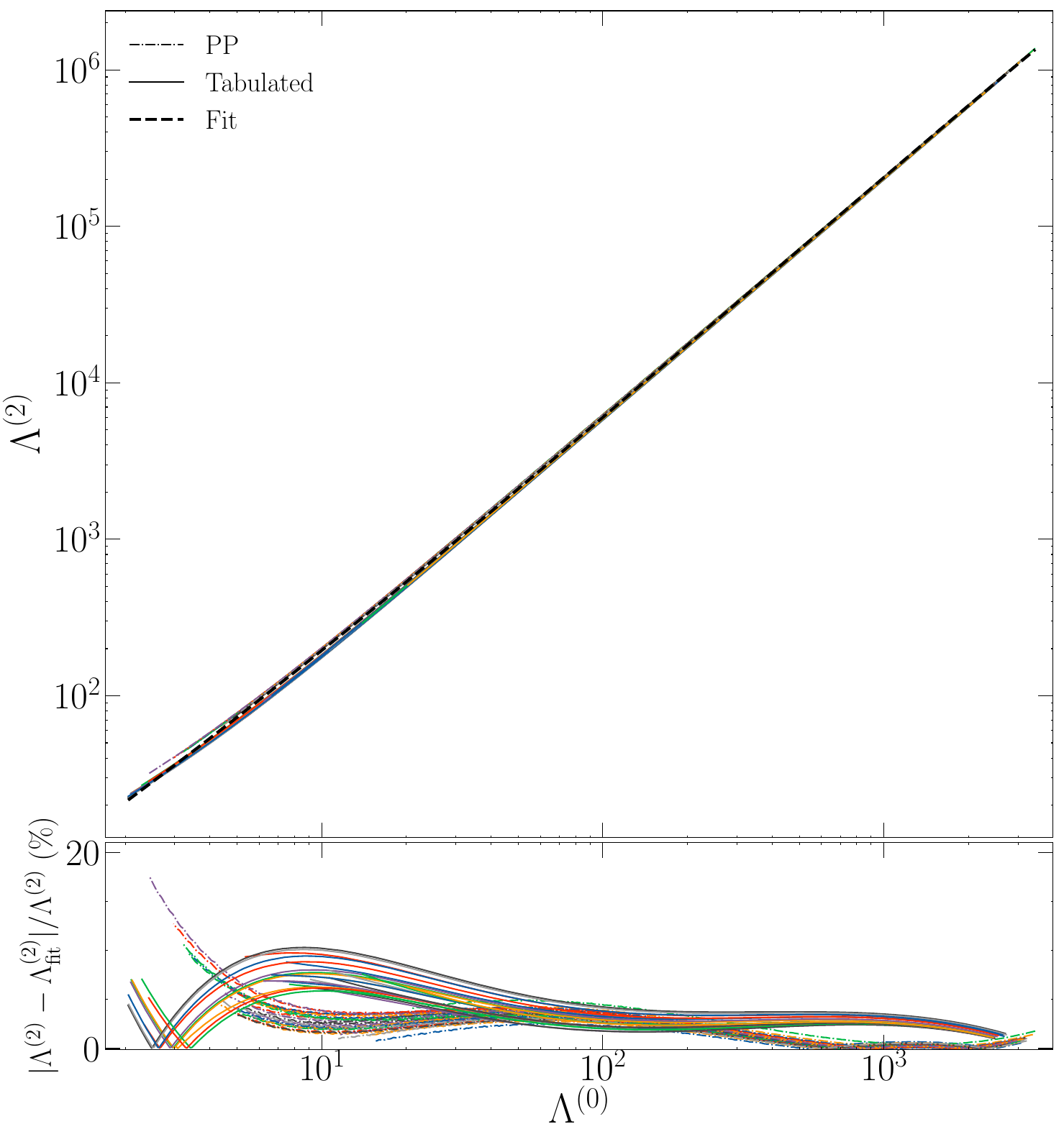}
    \caption{
    Relation between the dimensionless static tidal deformability $\Lambda^{(0)}$ and its second-order frequency correction $\Lambda^{(2)}$ for the complete set of EOS using the external solutions from~\cite{Pitre:2023xsr}. Compactness varies along each curve from the maximum allowed mass (left corner) down to $C=0.12$ (right corner).  
    \textbf{Top:} Relation $\Lambda^{(2)}$--$\Lambda^{(0)}$  in log-log scale. Here, solid lines correspond to the tabulated EOS, dash-dotted lines to the PP EOS set, and the thick dashed line to the fit relation~\eqref{eq:lambda_fit}. 
    \textbf{Bottom:} Fractional deviation from the best fit, expressed in percent. 
    }
    \label{fig:lambda_relation_dif_norm}
\end{figure}
Let us begin by describing the matching done when integrating $H^{(0)}(r)$. Because the differential equation describing $H^{(0)}(r)$ is homogeneous, matching is typically not required. In the calculation of the static Love number $k_{2}^{(0)}$, the scale $h^{(0)}_0$ cancels out. However, as discussed in Sec.~\ref{sec:Nearthecenter}, the constant $h^{(0)}_0$ reappears in the higher-order series expansions for $W^{(0)}(r)$, $V^{(0)}(r)$, and $H^{(2)}(r)$, and thus, we must find it. To determine $h^{(0)}_0$, we match the interior and exterior solutions at the surface. We impose continuity on both $H^{(0)}(r)$ and $H^{(0)'}(r)$ at $r=R$. Defining $\mathbf{Y}=\{H,H'\}$, we obtain: 
\begin{equation}
    h^{(0)}_0 \mathbf{Y}_\text{int}(R) = \mathbf{Y}_\text{ext}(R) 
\end{equation}
from which we obtain the usual expression for $k_{2}^{(0)}$
\begin{align}
    &\frac{1}{k_{2}^{(0)}} = \frac{60 z^2 \log (z)}{-(z-1)^5 z^2 } 
     \nonumber\\
     &+\frac{  5 \left[H^{(0)}_\text{int} \left(3 z^4-16 z^3+12 z^2+1\right)+H^{(0)'}_\text{int} R z \left(z^3-7 z^2-7 z+1\right)\right]}{-(z-1)^4 z^2 \left[H^{(0)}_\text{int} (z-1)+H^{(0)'}_\text{int} R z\right]}
\end{align}
and an expression for $h^{(0)}_0$
\begin{equation}
        h^{(0)}_0 = -\frac{10 }{z \left[H^{(0)}_\text{int} (z-1)+H^{(0)'}_\text{int} R z\right]} k_2,
\end{equation}
where $z$, $H^{(0)}_\text{int}$ and $H^{(0)'}_\text{int}$ are evaluated at $r=R$. These solutions allow us to proceed with the fluid perturbation calculations. 

Since the fluid perturbation functions are defined only within the stellar interior, we enforce boundary conditions at both the center and the surface of the star, allowing us to perform matching at an intermediate radius $r = r^* = R/2$. We start, again, by defining a vector representation of the functions $\mathbf{U} = \{W^{(0)},V^{(0)}\}$, and decompose the solution into homogeneous and particular components. We then integrate both components from the center (left integration), using an initial guess for $v_0$, and separately from the surface (right integration) with another initial guess for $v_R$. The full solutions at the matching point are constructed as 
\begin{equation}
    \mathbf{U^L} = \mathbf{U_\text{par}^{L}} + a_\mathbf{L} \mathbf{U_\text{hom}^{L}}, \quad
    \mathbf{U^R} = \mathbf{U_\text{par}^{R}} + a_\mathbf{R} \mathbf{U_\text{hom}^{R}}.
\end{equation}
Continuity is imposed by setting $\mathbf{U^L}(r^*)=\mathbf{U^R}(r^*)$, which determines $ a_\mathbf{L}$ and $a_\mathbf{R}$.

The above integration can be done for any set of initial guesses $v_0$ and $v_R$, but we find that the following iterative procedure yields higher numerical accuracy and stability. We start with an initial matching procedure, defined by the choices $v_0=1$ and $v_R=1$.
We then refine our guesses for $v_0$ and $v_R$ via
\begin{align}
    v_0 &= V_\text{par}^{L}(r=r_0) + a_\mathbf{L}V_\text{hom}^{L}(r=r_0) \nonumber \\ 
    v_R &= V_\text{par}^{R}(r=R) + a_\mathbf{R}V_\text{hom}^{R}(r=R),
\end{align}
and repeat the process. This procedure is carried out iteratively until the constants  $a_\mathbf{L}$ and $a_\mathbf{R}$ become sufficiently small, indicating convergence. Note that only the particular solutions need updated initial conditions, since the homogeneous solutions are fixed and do not depend on these values.

Lastly, we proceed to integrate the dynamical metric perturbation function $H^{(2)}$. For this step, we perform two sets of matching: one at the surface and another at the intermediate point $r^*$. Since we are interested only in calculating $k_{2}^{(2)}$ and not in the constants of integration $h^{(2)}_0$, $h^{(2)}_R$, and $h^{(2)'}_R$, the matching is done simultaneously. Let us begin by defining $\mathbf{Y^{(2)}} = \{H^{(2)}, H^{(2)'}\}$ as the vector encoding the quantities we shall match. At the center, there is only one free constant, $h^{(0)}_0$, and so we create a solution much like the one previously performed for the fluid integration. That is, we decompose $\mathbf{Y^{(2)L}}$ into a homogeneous and particular solution with an initial guess for $h^{(2)}_0$, in this case, we pick $h^{(2)}_0=1$. We then construct a full solution via 
\begin{equation}
    \mathbf{Y^{(2)L}} = \mathbf{Y^{(2)L}}_\text{par}+ b_\mathbf{L} \mathbf{Y^{(2)L}}_\text{hom},
\end{equation}
where $b_\mathbf{L}$ is a constant to be determined by matching.
At the surface, there are two free constants, which means we need to decompose the solution into three components: two homogeneous solutions and one particular solution. The two homogeneous solutions must be linearly independent to form a complete solution. For this reason, we choose two sets of initial guesses for $(h^{(2)}_R,h^{(2)'}_R)$, such as $\{1,0\}$ and $\{0,1\}$. The full solution is then constructed as
\begin{equation}
    \mathbf{Y^{(2)R}} = \mathbf{Y^{(2)R}}_\text{par}+ b_\mathbf{R}^1 \mathbf{Y^{(2)R,1}}_\text{hom}+ b_\mathbf{R}^2 \mathbf{Y^{(2)R,2}}_\text{hom}, 
\end{equation}
where $b_\mathbf{R}^{1,2}$ are constants to be determined by matching. Lastly, we have the exterior solution $\mathbf{Y^{(2)}}_\text{ext}$, which we write as the $\omega^2$ term in Eq.~\eqref{eq:exteriorHfull}, along with its analytic derivative. 

Next, we enforce continuity and differentiability both at the intermediate radius $r=r^*$ and at the surface $r=R$ by requiring that
\begin{equation}
    \mathbf{Y^{(2)L}}(r^*)=\mathbf{Y^{(2)R}}(r^*), \quad \mathbf{Y^{(2)R}}(R)=\mathbf{Y^{(2)}}_\text{ext}(R).
\end{equation}
These four equations can then be used to eliminate the free constants and determine $k_{2}^{(2)}$, completing the solution. This process is repeated iteratively, like in the fluid integration, until the constants $b_\mathbf{L}$, $ b_\mathbf{R}^2$, and $ b_\mathbf{R}^2$ are sufficiently small and $k_{2}^{(2)}$ is determined.

\section{Universal Relations using the external solutions from~\cite{Pitre:2023xsr}}
\label{Ap:Normalization}
The external solution used in~\cite{Pitre:2023xsr} differs from our solutions [Eqs.~\eqref{eq:Htide-ours} and \eqref{eq:Hmulti-ours}], see Sec. IV B of~\cite{HegadeKR:2024agt} for a detailed discussion.
Let us denote the external tidal and multipolar solutions used in~\cite{Pitre:2023xsr} with $\mathcal{H}_{\rm tide}$ and $\mathcal{H}_{\rm multi}$ respectively.
The explicit expressions for these functions are~\cite{Poisson:2020vap}
\begin{widetext}
\begin{align}
\mathcal{H}_{\rm tide}&= z + \frac{M^2 \omega ^2 }{630 (z-1)^2 z}\bigg\{-2520 (z-1)^2 z^2 \text{Li}_2(z)+107 z^6-210 z^6 \log (z)-860 z^5+2100 z^5 \log (z)+1538 z^4\nonumber\\
&-1260 z^4 \log ^2(z)-3570 z^4 \log (z)-2074 z^3+2520 z^3 \log ^2(z)+3989 z^2-1260 z^2 \log ^2(z)+3570 z^2 \log (z)\\&+6 (z-1)^2 \left(35 z^4-280 z^3+214 z^2+280 z-35\right) \log (1-z)-3430 z-2100 z \log (z)+210 \log (z)+70\bigg\}\nonumber
\end{align}
\end{widetext}
\begin{widetext}
\begin{align}
\mathcal{H}_{\rm multi}&= -\frac{5 \left(z^4-8 z^3+12 z^2 \log (z)+8 z-1\right)}{32 C^5 z} -\frac{M^2 \omega^2}{2016 C^5 (z-1)^2 z}\bigg\{30240 (z-1)^2 z^2 \text{Li}_3(z)+36 (z-1)^2 \text{Li}_2(z) \big[35 z^4\nonumber\\&-280 z^3+428 z^2-420 z^2 \log (z)+280 z-35\big]+857 z^6-630 z^6 \log ^2(z)-642 z^6 \log (1-z)+1260 z^6 \log (1-z) \log (z)
\nonumber\\&+642 z^6 \log (z)-7952 z^5+6300 z^5 \log ^2(z)+6420 z^5 \log (1-z)-12600 z^5 \log (1-z) \log (z)-5160 z^5 \log (z)\nonumber\\&-30240 z^4 \zeta (3)-2568 \pi ^2 z^4+19163 z^4-2520 z^4 \log ^3(z)-10710 z^4 \log ^2(z)-10914 z^4 \log (1-z)\\&+29124 z^4 \log (1-z) \log (z)+9228 z^4 \log (z)+60480 z^3 \zeta (3)+5136 \pi ^2 z^3-25900 z^3+5040 z^3 \log ^3(z)\nonumber\\&-15408 z^3 \log (1-z) \log (z)-12444 z^3 \log (z)-30240 z^2 \zeta (3)-2568 \pi ^2 z^2+22487 z^2-2520 z^2 \log ^3(z)\nonumber\\&+10710 z^2 \log ^2(z)+10914 z^2 \log (1-z)-13716 z^2 \log (1-z) \log (z)+23934 z^2 \log (z)-5348 z-6300 z \log ^2(z)\nonumber\\&+630 \log ^2(z)-6420 z \log (1-z)+12600 z \log (1-z) \log (z)-20580 z \log (z)+642 \log (1-z)\nonumber\\&-1260 \log (1-z) \log (z)+420 \log (z)-3307\bigg\}\nonumber
\end{align}
\end{widetext}

\begin{figure}[h!]
    \centering
    \includegraphics[width=\linewidth]{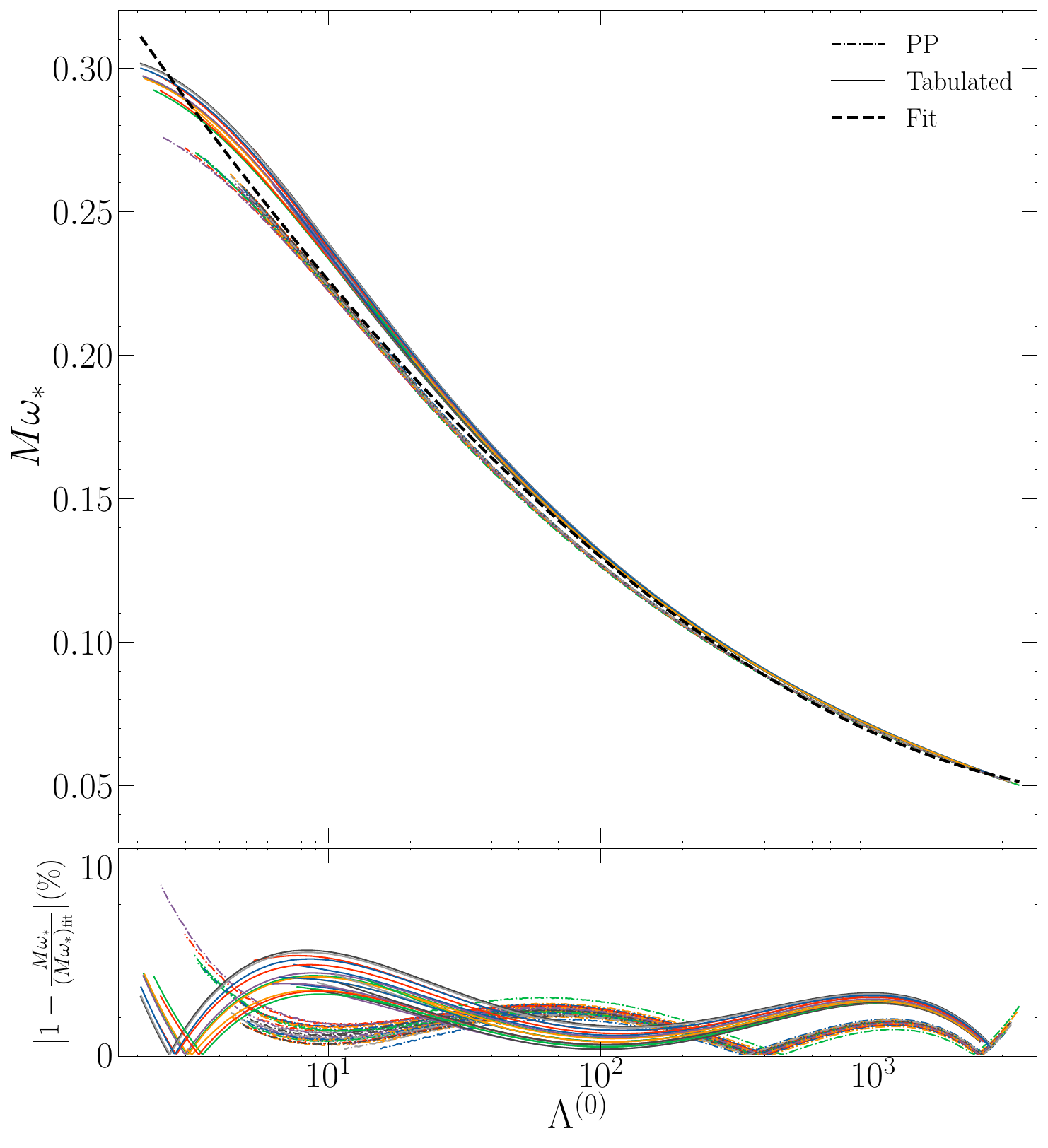}
    \caption{
    Relation between the dimensionless static tidal deformability $\Lambda^{(0)}$ and the effective dimensionless frequency $M \omega_*$ for the EOSs studied in this paper using the external solutions from~\cite{Pitre:2023xsr}. Compactness varies along each curve from the maximum allowed mass (left corner) down to $C=0.12$ (right corner).  
    \textbf{Top:} Relation $\Lambda^{(0)}$--$M \omega_*$. Here, solid lines correspond to the tabulated EOS, dash-dotted lines to the PP set EOS, and the thick dashed line to the fit relation~\eqref{eq:omega_fit}. 
    \textbf{Bottom:} Fractional deviation from the best fit, expressed in percent. 
    }
    \label{fig:lambda_y_relation_dif_norm}
\end{figure}

Using the above solutions, Fig.~\ref{fig:lambda_relation_dif_norm} shows the $\Lambda^{(0)}$--$\Lambda^{(2)}$ universal relation. The best-fit parameters in this case are given by $ (a_0,a_1,a_2,a_3) = (2.140, 1.267, 4.651\times 10^{-2}, -2.652\times 10^{-3})$. For the $\Lambda^{(0)}$--$M\omega_*$ relation, see Fig.~\ref{fig:lambda_y_relation_dif_norm}. The best-fit parameters for the curves in Fig.~\ref{fig:lambda_y_relation_dif_norm} are $(b_0, b_1, b_2, b_3) = (0.3538, -6.176 \times 10^{-2}, 2.614\times 10^{-3}, 5.044\times 10^{-5})$.

\bibliographystyle{apsrev4-1}

\end{document}